\documentclass[preprint]{aastex631}

\usepackage{url}
\usepackage{amsmath}
\usepackage{amsfonts}
\usepackage{amssymb}
\usepackage{multirow}
\usepackage{comment}
\usepackage{dsfont}
\usepackage{graphicx}
\graphicspath{{bd_plot/}}

\def \apjs {{\rm {ApJS}}}

\def \aap {{\rm {A\&A}}}

\def \apjl{\rm {ApJ}}

\def \deg{$^\circ$}

\begin{document}

\title{Tianyu: search for the second solar system {and explore the dynamic universe}}

\author{Fabo Feng}
\altaffiliation{ffeng@sjtu.edu.cn}
\affiliation{Tsung-Dao Lee Institute, Shanghai Jiao Tong University, Shengrong Road 520, Shanghai 201210}
\affiliation{School of Physics and Astronomy, Shanghai Jiao Tong University, 800 Dongchuan Road, Shanghai 200240}

\author{Yicheng Rui}
\altaffiliation{ruiyicheng@sjtu.edu.cn}
\affiliation{Tsung-Dao Lee Institute, Shanghai Jiao Tong University, Shengrong Road 520, Shanghai 201210}

\author{Zhimao Du}
\affiliation{Shanghai Astronomy Museum (Branch of Shanghai Science and Technology Museum), Shanghai 200127}

\author{Qing Lin}
\affiliation{Shanghai Astronomy Museum (Branch of Shanghai Science and Technology Museum), Shanghai 200127}

\author{Congcong Zhang}
\affiliation{Shanghai Astronomical Observatory, Chinese Academy of Sciences, 80 Nandan Road, Shanghai 200030}

\author{Dan Zhou}
\affiliation{Shanghai Astronomical Observatory, Chinese Academy of Sciences, 80 Nandan Road, Shanghai 200030}

\author{Kaiming Cui}
\affiliation{Tsung-Dao Lee Institute, Shanghai Jiao Tong University, Shengrong Road 520, Shanghai 201210}

\author{Masahiro Ogihara}
\affiliation{Tsung-Dao Lee Institute, Shanghai Jiao Tong University, Shengrong Road 520, Shanghai 201210}
\affiliation{School of Physics and Astronomy, Shanghai Jiao Tong University, 800 Dongchuan Road, Shanghai 200240}

\author{Ming Yang}
\affiliation{College of Surveying and Geo-Informatics, Tongji University, Shanghai 20009}

\author{Jie Lin}
\affiliation{CAS Key laboratory for Research in Galaxies and Cosmology, Department of Astronomy, University of Science and Technology of China, Hefei 230026}
\affiliation{School of Astronomy and Space Sciences, University of Science and Technology of China, Hefei 230026}

\author{Yongzhi Cai}
\affiliation{Key Laboratory for the Structure and Evolution of Celestial Objects, Chinese Academy of Sciences, Kunming 650216}
\affiliation{Yunnan Observatories, Chinese Academy of Sciences, Kunming 650216}
\affiliation{International Centre of Supernovae, Yunnan Key Laboratory, Kunming 650216}

\author{Taozhi Yang}
\affiliation{School of Physics, Xi'an Jiaotong University, Xi'an 710049}

\author{Xiaoying Pang}
\affiliation{Department of Physics, Xi'an Jiaotong-Liverpool University, 111 Ren’ai Road, Dushu Lake Science and Education Innovation District, Suzhou 215123}
\affiliation{Shanghai Key Laboratory for Astrophysics, Shanghai Normal University, 100 Guilin Road, Shanghai 200234}

\author{Mingjie Jian}
\affiliation{Department of Astronomy, Stockholm University, AlbaNova University Center, Roslagstullsbacken 21, 114 21 Stockholm, Sweden}

\author{Wenxiong Li}
\affiliation{National Astronomical Observatories, Chinese Academy of Sciences, Beijing 100101}

\author{Hengxiao Guo}
\affiliation{Shanghai Astronomical Observatory, Chinese Academy of Sciences, 80 Nandan Road, Shanghai 200030}

\author{Xian Shi}
\affiliation{Shanghai Astronomical Observatory, Chinese Academy of Sciences, 80 Nandan Road, Shanghai 200030}

\author{Jianchun Shi}
\affiliation{Shanghai Astronomical Observatory, Chinese Academy of Sciences, 80 Nandan Road, Shanghai 200030}

\author{Jianyang Li}
\affiliation{School of Atmospheric Sciences, Sun Yat-sen University, Zhuhai 519082}

\author{Kangrou Guo}
\affiliation{Tsung-Dao Lee Institute, Shanghai Jiao Tong University, Shengrong Road 520, Shanghai 201210}

\author{Song Yao}
\affiliation{Shanghai Astronomy Museum (Branch of Shanghai Science and Technology Museum), Shanghai 200127}

\author{Aming Chen}
\affiliation{Tsung-Dao Lee Institute, Shanghai Jiao Tong University, Shengrong Road 520, Shanghai 201210}

\author{Peng Jia}
\affiliation{College of Electronic Information and Optical Engineering, Taiyuan University of Technology, Taiyuan 030024}

\author{Xianyu Tan}
\affiliation{Tsung-Dao Lee Institute, Shanghai Jiao Tong University, Shengrong Road 520, Shanghai 201210}
\affiliation{School of Physics and Astronomy, Shanghai Jiao Tong University, 800 Dongchuan Road, Shanghai 200240}

\author{James S. Jenkins}
\affiliation{Instituto de Estudios Astrof\'isicos, Facultad de Ingenier\'ia y Ciencias, Universidad Diego Portales, Av. Ej\'ercito 441, Santiago, Chile}
\affiliation{Centro de Astrof\'isica y Tecnolog\'ias Afines (CATA), Casilla 36-D, Santiago, Chile}

\author{Hongxuan Jiang}
\affiliation{Tsung-Dao Lee Institute, Shanghai Jiao Tong University, Shengrong Road 520, Shanghai 201210}

\author{Mingyuan Zhang}
\affiliation{Tsung-Dao Lee Institute, Shanghai Jiao Tong University, Shengrong Road 520, Shanghai 201210}

\author{Kexin Li}
\affiliation{Tsung-Dao Lee Institute, Shanghai Jiao Tong University, Shengrong Road 520, Shanghai 201210}
\affiliation{Shandong Key Laboratory of Optical Astronomy and Solar-Terrestrial Environment, School of Space Science and Physics, Institute of Space Sciences, Shandong University, Weihai, Shandong 264209}

\author{Guangyao Xiao}
\affiliation{Tsung-Dao Lee Institute, Shanghai Jiao Tong University, Shengrong Road 520, Shanghai 201210}

\author{Shuyue Zheng}
\affiliation{Tsung-Dao Lee Institute, Shanghai Jiao Tong University, Shengrong Road 520, Shanghai 201210}

\author{Yifan Xuan}
\affiliation{Tsung-Dao Lee Institute, Shanghai Jiao Tong University, Shengrong Road 520, Shanghai 201210}

\author{Jie Zheng}
\affiliation{CAS Key Laboratory of Optical Astronomy, National Astronomical Observatories, Chinese Academy of Sciences, Beijing 100101}

\author{Min He}
\affiliation{CAS Key Laboratory of Optical Astronomy, National Astronomical Observatories, Chinese Academy of Sciences, Beijing 100101}

\author{Hugh R.A. Jones}
\affiliation{Centre for Astrophysics Research, University of Hertfordshire, Hatfield, Hertfordshire, AL10 9AB, UK}

\author{Cuiying Song}
\affiliation{Physics Department and Tsinghua Center for Astrophysics, Tsinghua University, Beijing 100084, China}


\begin{abstract}
Giant planets like Jupiter and Saturn, play important roles in the formation and habitability of Earth-like planets. The detection of solar system analogs that have multiple cold giant planets is essential for our understanding of planet habitability and planet formation. Although transit surveys such as Kepler and Transiting Exoplanet Survey Satellite (TESS) have discovered thousands of exoplanets, these missions are not sensitive to long period planets due to their limited observation baseline. The Tianyu project, comprising two 1-meter telescopes (Tianyu-I and II), is designed to detect transiting cold giant planets in order to find solar system analogs. Featuring a large field of view and equipped with a high-speed CMOS camera, Tianyu-I will perform a high-precision photometric survey of about 100 million stars, measuring light curves at hour-long cadence. The candidates found by Tianyu-I will be confirmed by Tianyu-II and other surveys and follow-up facilities through multi-band photometry, spectroscopy, and high resolution imaging. Tianyu telescopes will be situated at an elevation about 4000 meters in Lenghu, chosen as the premier observation site in China. With a photometric precision of 0.1\% for stars with $V<14$\,mag, and 1\% for stars with $V<18$\,mag, Tianyu is expected to find more than 300 transiting exoplanets, including about 12 cold giant planets, over five years. Assuming coplanarity for solar system analogs and an occurrence rate of 10\% for Earth twins, a five-year survey of Tianyu would discover 1-2 solar system analogs, which could be confirmed by Earth-hunting missions such as Earth 2.0. Moreover, Tianyu is also designed for non-exoplanetary exploration, incorporating multiple survey modes covering timescales from sub-seconds to months, with a particular emphasis on events occurring within the sub-second to hour range. It excels in observing areas such as infant supernovae, rare variable stars and binaries, tidal disruption events, Be stars, cometary activities, and interstellar objects. These discoveries not only enhance our comprehension of the universe but also offer compelling opportunities for public engagement in scientific exploration.
\end{abstract}


\section{Introduction}\label{sec:intro}

Are we alone in the universe? Is our Earth or Solar System rare? To answer these and other related questions scientifically, astronomers have detected more than 5000 exoplanets since the discovery of the first exoplanet orbiting around a solar analog \citep{mayor95}. 
Solar System (SS) formation is shaped by events like nearby supernovae \citep{1977Icar...30..447C}, Jupiter and Saturn migration \citep{2005Natur.435..459T}, moon-forming impacts\citep{stevenson87}, and the Late Heavy Bombardment\citep{gomes05}, influencing the occurrence of Solar System analogs. The Solar System architecture, particularly with multiple cold giants, plays a key role in Earth's habitability, as seen in Jupiter's comet-scattering during early formation. The discovery of Jupiter and Neptune-like cold giants (CGs) is essential for grasping their formation and evolution \citep{raymond17,piani20}. 

Specifically, a solar system analog is defined as a solar analog consist of at least one Earth twin and a CG \citep{yahalomi23}. A solar analog has a mass and radius of 0.75-1.25$M_\odot$ and 0.75-1.25$R_\odot$, respectively. An Earth twin is defined as a planet with a mass and radius of 0.5-2$M_\oplus$ and 0.75-1.25$R_\oplus$, respectively. 
The CGs, analogous to Jupiter, Saturn, Uranus, and Neptune, have masses ranging from 10 to 1000$M_\oplus$ (or 0.03 to 3 $M_{\rm Jup}$) and orbital periods exceeding 3 years\citep[e.g.,][]{rowan16}. The lower limit of the orbital period corresponds to the inner boundary of the snowline region, which is at about 2\,au \citep{mulders15}. While several projects, including Earth 2.0 \citep{ge22}, CHES\citep{ji22}, and THE \citep{hall18}, are planning to detect Earth-like planets, a crucial aspect of exploring solar system analogs{, a}nother key aspect is the detection of CGs in these systems. 

\begin{figure}
	\centering
	\includegraphics[width = 0.8\linewidth]{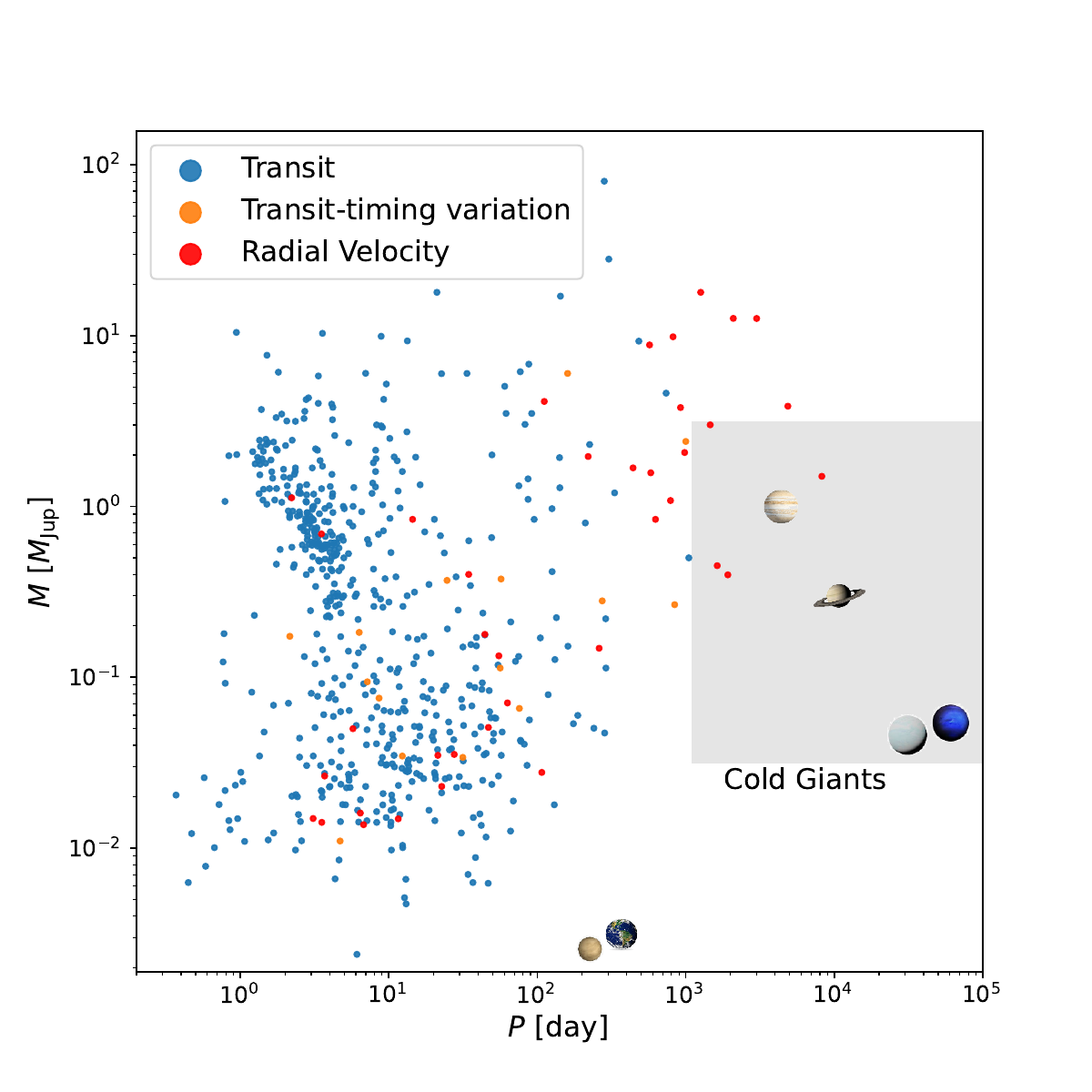}
	\caption{Exoplanets confirmed with absolute mass around solar analogs. Dot colors denote the discovery method, while the shaded region signifies the parameter space occupied by CGs. The exoplanet parameters are obtained {f}rom NASA Exoplanet Archive \protect\url{https://exoplanetarchive.ipac.caltech.edu/}. Venus, Earth, Jupiter, Saturn, Uranus and Neptune are shown for reference.}
	\label{fig:current_solar_like}
\end{figure}

As shown in Fig. \ref{fig:current_solar_like}, the current detection methods are not sensitive to CGs due to many reasons. First, as the most productive method\citep{2011ApJ...736...19B}, the transit method, is not sensitive to long period planets due to the short total observation time (typically less than 4 years) of individual surveys. Second, although sensitive to long period planets, the radial velocity method cannot determine the absolute dynamical mass of a planet due to degeneracy between mass and inclination\citep{2010exop.book...27L,2018ARA&A..56..175D}. Third, although the microlensing method\citep{1964MNRAS.128..295R} is sensitive to CGs on wide orbits, the mass and orbital elements of these CGs are typically poorly determined partly due to lack of follow-up observations. Finally, the imaging method is only sensitive to a limited number of young Jupiters and thus cannot survey CGs as old as those in our solar system.

Acknowledging the limitations of individual methods, the synergy of radial velocity and astrometry emerges as a powerful tool for CG detection, utilizing Gaia's rich and high-precision astrometric data \citep[e.g.][]{gaia18}. However, the rare occurrence of transiting Earth twins in systems identified through this combined approach makes their transit detection unlikely. In contrast, the identification of transiting CGs could significantly increase the chances of detecting transits of Earth twins in the same system, especially if the system is as coplanar as the solar system. Furthermore, the characterization of transiting CGs could be enhanced by observing their transmission spectra during the transit.


Transit missions like Kepler \citep{borucki10} and Transiting Exoplanet Survey Satellite (TESS) \citep{ricker14} focus on short-period transiting planets due to their limited lifespan. To identify long-period transits caused by CGs, the Tianyu\footnote{Tianyu derives its name from the renowned words of the esteemed Nobel laureate Tsung-Dao Lee, encapsulating the essence as "comprehend physics through the language of the heavens".} project deploys two 1-meter telescopes (Tianyu-I and Tianyu-II). Covering the Kepler and TESS continuous viewing zone, Tianyu-I will conduct a photometric survey of around 10 million stars with a 1-hour sampling frequency. Tianyu-I, with a 10-square-degree field of view (FOV), exhibits an etendue (survey efficiency) about one quarter of the Zwicky Transient Facility (ZTF; \citep{bellm19}). While Tianyu-I performs large-scale surveys, Tianyu-II is tailored for follow-up observations of candidates identified by Tianyu-I.

With a combined decade-long observation baseline from ground transit surveys, Kepler, TESS, Tianyu, and upcoming missions like Earth 2.0 \citep{ge22} and Plato \citep{ragazzoni16}, Tianyu is well-positioned to identify long-period transits. Candidates identified by Tianyu must exhibit at least two transits in its data or in other surveys. Further confirmation will rely on the high resolution spectrometer installed on the upcoming Jiao Tong University Spectroscopic Telescope (JUST; \citep{justteam}) and the Gaia astrometric data\footnote{\url{https://www.cosmos.esa.int/web/gaia/release}}. Although radial velocity and astrometric methods are common for CG detection \citep{wittenmyer20, kervella19, feng22}, Tianyu's identification of transiting CGs presents ideal targets for the James Webb Space Telescope (JWST; \cite{rigby23}) to pioneer atmospheric characterization through transmission spectroscopy.

Tianyu's survey strategy is strategically designed to enhance contributions to various non-exoplanetary scientific areas, including variable stars, solar system small bodies, time-domain, and multi-messenger science. The survey time is divided into three modes: short-cadence (subsecond to 1 hour), medium (1 hour to 1 week), and long-cadence (1 week to 1 month). This multi-mode approach improves the likelihood of capturing transients occurring over various timescales, from sub-seconds to hours. Examples include near-Earth objects (NEOs), trans-Neptune object (TNO) occultations, infant supernovae, fast blue optical transients, and optical counterparts of gamma-ray bursts (GRBs) and fast radio bursts (FRBs). Tianyu's high-cadence mode is also adept at detecting rare variable stars on minute-to-hour timescales, such as eclipsing binary white dwarfs, AM CVn, and blue large-amplitude pulsators \citep{lin23}. The low-cadence data will be instrumental in identifying `Oumuamua-like objects, cometary activity, and Be stars in star clusters.

The paper follows a structured format. Section \ref{sec:goals} introduces the scientific goals of the Tianyu project. Section \ref{sec:instr} details the observation site, telescopes, and instruments, and section \ref{sec:strategy} describes the survey strategy and data processing. The anticipated photometric and timing precision is detailed in section \ref{sec:precision}, while section \ref{sec:confirmation} outlines the strategy for validating transits identified by Tianyu. For a comprehensive understanding of the primary scientific results, refer to section \ref{sec:yield}. Finally, section \ref{sec:conclusion} provides the concluding remarks.

\section{Scientific Goals}\label{sec:goals}

\subsection{Exoplanetary science}
\subsubsection{Effects of giant planets on planet formation and evolution}
In the Solar system, CGs (i.e., Jupiter, Saturn, Uranus and Neptune) are thought to have influenced the orbits and properties of the inner Solar system in the past.
\begin{itemize}
	\item During the runaway gas accretion of giant planets, the giant planets scatter surrounding planetesimals and hence material mixing occurs \citep{2017Icar..297..134R,2020PSJ.....1...45C,ogihara2022}. Because the Earth's ocean is considered to originate from carbonaceous chondrites \citep{2008Icar..194...42G,2012E&PSL.313...56M}, it is possible that this process delivered the Earth's water.
	\item During the orbital migration of giant planets, the orbits of terrestrial planets can also be significantly affected. The Grand Tack model, which postulates that Jupiter and Saturn underwent large-scale migration, may have determined the outer boundary of the orbital distribution of the terrestrial planets \citep{2011Natur.475..206W}.
	Water transport to Earth may also have occurred during/after the Grand Tack migration \citep{2014Icar..239...74O}.
	It is also possible that giant planets formed quite distant orbits and migrated to their current orbits of Jupiter and Saturn \citep{2017Icar..297..134R,2022SciA....8M3045L}.
	\item Secular resonances associated with giant planets alter the inner Solar system. Secular resonances occur when the system's eigenfrequencies, determined mainly by giant planets, and the eigenfrequencies of other bodies are close to each other. This can significantly change the orbits of the bodies \citep{1961mcm..book.....B}. Since the position of the secular resonance depends on the potential of the protoplanetary gas disk, the position of the secular resonance moves with the dissipation of disk gas. This strongly affects the orbits of the inner planets \citep{1976Icar...28..441W,2000AJ....119.1480N}.
	\item The orbital instability of giant planets after disk dissipation can also affect the properties of planetary systems. Giant planets can undergo orbital instabilities that can alter the orbital configurations of giant planets as well as the orbits of lower-mass planets and small bodies. The Nice model suggests that this instability occurred 500--700 million years after disk dissipation \citep{2005Natur.435..466G}, and another model suggests that it occurred immediately after disk dissipation \citep{2018Icar..311..340C}.
\end{itemize}

These physical phenomena caused by giant planets depend on the mass, orbital configuration, and formation time of giant planets. In other words, information about the orbits and properties of giant planets is essential to discuss the formation and evolution of planetary systems.

Through the strategic optimization of survey methods to extend the observational baseline of transits, the Tianyu project aims to discover planetary systems featuring a combination of close-in planets and CGs (see section \ref{sec:exoplanet_yield} for the expected yield of such systems). This endeavor promises valuable insights into the underlying physical processes at play within each planetary system. In particular, for systems with CGs at around the semimajor axis of $a=5-10\,{\rm au}$ and terrestrial planets at around $a=1\,{\rm au}$, it is possible to study the specific effects of each physical phenomenon on the inner terrestrial planets. Such a study will clarify the formation and evolution of each exoplanet system, and also lead to a discussion of how much of each physical process occurred in the Solar system.

\subsubsection{Constraints on orbital migration from mass distribution and mean-motion resonances of giant planets}

Recent planet formation theory shows that orbital migration is a major factor in determining the orbital distribution of planets. Planetary orbital migration can be divided into type I migration for low-mass planets \citep{1979ApJ...233..857G} and type II migration for giant planets that open density gaps in the protoplanetary disk \citep{1986ApJ...309..846L}. The speed and direction of orbital migration can vary greatly depending on the properties of the planet and the protoplanetary disk for both type I migration \citep{2010A&A...523A..30B,2011MNRAS.410..293P,2014MNRAS.444.2031P,2015Natur.520...63B,2015A&A...584L...1O,2016ApJ...832..105F,2019MNRAS.484..728M} and type II migration \citep{2014ApJ...792L..10D,2015A&A...574A..52D,2018ApJ...861..140K}.
However, it is not easy to constrain how fast planets actually migrate in each planetary system.

Meanwhile, recent observations of exoplanets have provided a lot of information about super-Earths and sub-Neptunes in close-in orbits. Theoretical studies using this information suggest that many of these planets would not have experienced large-scale orbital migration \citep{2017A&A...607A..67M,2018A&A...615A..63O}.
In the same way, once the Tianyu project obtains information on the orbits and masses of CGs, constraints can be placed on the orbital migration rates of type II migration that CGs experienced.

One important piece of information is the planetary mass distribution of CGs. If multiple CGs are discovered in a system, their mass distribution could constrain the orbital migration. If the inner CGs have larger masses, they may have experienced some degree of orbital migration \citep{2019A&A...622A.202J,2019A&A...623A..88B}. On the other hand, if the mass distribution in the system is nearly constant, it is likely that CGs have not migrated or have undergone convergent migration \citep{2020MNRAS.496.3314W}.

Another important piece of information is mean-motion resonances (MMRs). Mean-motion resonances can play an important role in planet formation and evolution. For example, the Nice model and the Grand Tack model assume that giant planets are captured in MMRs.
It is known that for planets to be captured in MMRs, they must undergo convergent migration at a certain migration speed \citep{2011MNRAS.413..554M,2013ApJ...775...34O}.
Therefore, the discovery of a pair of CGs  in MMRs by the Tianyu project would provide useful information about the rate of orbital migration that CGs experienced during their formation. Assuming an occurrence rate of 10\% for CGs, Tianyu is expected to find about 1-2 systems with double CGs (see section \ref{sec:exoplanet_yield}).
In addition to constraining orbital migration rates, information about MMRs is of great importance for the origin of planetary systems. While several systems with giant planets in MMRs have been discovered so far, the discovery of even one system with Jupiter and Saturn analogs in MMRs would be a significant finding and an important target for future study.
In fact, the TRAPPIST-1 system in MMRs has attracted much attention and has been the subject of various studies \citep{2021ApJ...907...81L,2022MNRAS.511.3814H}.

\subsubsection{Coexisting system of Jupiters and Neptunes}

The Jupiters of the Solar system (Jupiter and Saturn) have a large amount of gas ($\gtrsim 100\,M_\oplus$), while the Neptunes(Neptune and Uranus) have not accreted much gas ($\sim 10\,M_\oplus$). Solar system formation theory suggests that this difference is due to different growth timescales of planetary cores \citep[e.g.,][]{safronov1969evolution}.
Jupiter's core grew rapidly and started runaway gas accretion well before the disk gas dispersed, while for Saturn, the disk dissipated during the runaway gas accretion and the gas accretion onto the core was quenched in the middle.
Uranus and Neptune did not have time for runaway gas accretion to occur because their cores grew at about the same time as the disk dissipated \citep[e.g.,][]{Pollack_et_al_1996}.
In this way, when Jupiters and Neptunes coexist in a system, it is possible to place constraints on the timing relative to the disk lifetime at which runaway gas accretion occurred for those planets.

It is expected to find a few systems with Jupiters and Neptunes coexisting from the Tianyu project (see section \ref{sec:yield}). In this case, we can put a constraint on the timing of the growth of these planets and expect to obtain information on how they formed (i.e., planetesimal accretion, pebble accretion, gravitational instability).

\subsubsection{Occurrence rates of CGs}

Occurrence rates of exoplanets have recently been estimated \citep{Howard2010,Mayor2011,Fressin2013,Zhu2018}.
The occurrence rates of hot super-Earths and hot/warm Jupiters around solar-type stars are now relatively well understood, allowing for various discussions on their origins \citep[e.g.,][]{Wright2012,Wang2015}.
However, there are only some limited estimates from radial velocity surveys for CGs with their orbits at about $a = 1-10\,{\rm au}$ \citep{Cumming2008,Fernandes2019,Fulton2021}, and they are not yet restricted enough to be compared to those for super-Earths and hot Jupiters. In addition, the frequency of CGs around M-dwarfs are far less understood compared with those around solar-type stars due to paucity of detection \citep{Bonfils2013,Montet2014,Bowler2015,Clanton2016,Sabotta2021,Pinamonti2022}. 

In addition to radial velocity and astrometry \citep{feng22}, the occurrence rate of planets in wide orbits at distances greater than $a = 10\,{\rm au}$ has recently been estimated by imaging \citep{Nielsen2019,Vigan2021} and microlensing \citep{Poleski2021}.
In terms of transit survey, the Tianyu project would provide additional information on the occurrence rate of CGs, which together with the previous estimates may further constrain the occurrence of CGs. 
These allow discussion of occurrence rates over a wide range of orbits from hot planets to warm, cold, and extremely cold planets, providing valuable evidence for planet formation and migration history.


\subsubsection{Finding exoplanets in unexplored parameter space of atmospheric characterization}
Atmospheric characterization reveals the surface and atmospheric compositions and thermal structures, climate, and spin states of exoplanets, therefore providing important information on the formation and evolution of planetary systems as well as their potential habitability. Partly due to current limitations of our instrumental sensitivities, atmospheric characterization so far has mainly focused on two extreme populations of exoplanets:  1) short-period exoplanets that are heavily irradiated by their host stars and 2) extremely distant or free-floating self-luminous giant planets \citep{Madhusudhan2019,showman2020,guillot2022}. The new commissioning of JWST provides opportunities to characterize exoplanets with effective temperatures substantially lower than 1000 K (e.g., \citealp{kempton2023,Madhusudhan2023}); future ground-based instruments mounted on 30-meter class telescopes are expected to characterize much more exoplanets with moderate orbital distances and atmospheric temperatures (e.g., \citealp{ELT2023}). 

While current transit surveys, as exemplified by \citep{kipping16}, have detected only 1-2 transiting CGs, Tianyu is poised to identify $\sim$10 transiting CGs throughout its five-year survey duration (refer to section \ref{sec:exoplanet_yield}). This significant increase will expand the existing sample by an order of magnitude. Hence Tianyu project will be optimized to search for a much larger yet uncharacterized population of exoplanets with atmospheric temperatures much lower than 1000\,K and with orbital periods from tens of days to Jovian-like periods, aiding in finding optimal exoplanets in support of the ongoing and forthcoming exoplanet characterization programs. Tianyu's wide FOV and optimized target selection strategy in conjunction with other exoplanet searching missions are unique among the ground-based programs, which would be an important contribution to broadening planet population in terms of orbital distances amenable for atmospheric characterization. 

\subsubsection{Characterization of Exoplanet Systems via {transit-timing variation (TTV) }}
In a system with just one planet, we would expect to see transits happen with clockwork regularity. Nonetheless, the presence of additional planets can lead to gravitational interactions that disrupt this regularity. The gravitational pull from these neighboring planets can subtly alter the speed of the transiting planet as it orbits its star, either hastening or delaying its path. Consequently, the intervals between consecutive transits can vary, leading to TTV\citep{HM05, Ago05}. The TTV technique has played an important role in confirming multi-planet systems \citep[e.g.][]{For11,Ming13,Xie14,Hol16}. Moreover, TTV analysis provides insights into the configuration and dynamics of planetary systems, including resonances between planetary orbits. Most importantly, TTVs can be used to determine the masses of the planets \citep{Lit12,Had17}, which is crucial for understanding their compositions and potential habitability. The {\it Kepler} space mission, with its long-cadence of approximately 30 minutes, has identified over a hundred exoplanets using TTV technique. 

{The tidal interaction between stars and their planets are also important for understanding the long-term evolution of planetary systems \citep{Rao2021}. Transit-timing data are important for discovering systems with tidal orbital decay \citep{Patra_2020}. So far, WASP 12-b is the only planetary system that have confirmed tidal orbital decay \citep{Turner_2021}. One difficulty of confirming tidal orbital decays is their degeneracy with other effects which would lead to  signals, e.g. apsidal precession and Romer delay caused by unknown companion\citep{Harre23}. Huge amounts of transit-timing data combined with radial velocity measurements are needed to rule out these effects \citep{Yee_2020}.}

{With a sampling frequency ranging from sub-seconds to hours, Tianyu data will significantly enhance precision and extend the observation baseline of TTV signals. This capability is crucial for unraveling the diverse factors contributing to TTV phenomena. A simulation is conducted to show the potential of Tianyu to improve timing precision (see section \ref{subsec:timing_precision}).} By conducting joint analyses of these TTV data alongside Gaia astrometry (similar to the combined radial velocity and astrometric analyses done in \citep{feng22}), we anticipate discovering numerous non-transiting CGs. These findings hold the potential to play a crucial role in understanding the formation and evolution of close-in transiting planets.

\subsubsection{Hot Jupiters with Cold Neptunes}
Less than 10\% percent of Sun-like stars host gas giant planets orbiting within 1\,au of their host stars \citep{Fressin13}, 
and from these there exists three general populations, the hot Jupiters, the eccentric giants, and the circular giants. Since 
these giant planets are believed to have formed beyond the snow line at distances of $\sim$5~AU, they must have migrated from 
their original positions into their current orbits.  Migrating giant planets are expected to dynamically scatter any inner worlds 
as they pass through the inner system, and if they migrate by a high-eccentricity process, such as Kozai-Lidov secular 
interactions with long-period more massive companions \citep[e.g.][]{vick19}, we might not expect additional smaller planets to be present out 
to a few AUs in these systems.  Recently, radial-velocity analyses may have uncovered a new mechanism to form these eccentric
giants, which come along with additional Neptune-sized planets being found exterior to the eccentric giant. Rubenstein et al.
(2024, submitted) have discovered two super-Neptunes orbiting exterior to the previously known and moderately eccentric giant HD208487b.  Since the three planets are close to a Laplace chain resonance, they propose a model that begins with four super-Neptunes in resonance, the so called ``peas in a pod''  configuration \citep{weiss18}, and ends with the inner two planets colliding 
to create the eccentric gas giant, and leaving the two outer planets undisturbed. 

By searching for CGs around stars hosting known hot Jupiters, the Tianyu project may find about 10 such types of systems (see section \ref{sec:exoplanet_yield}), testing what fraction of eccentric inner gas giants have outer smaller companions, and therefore providing a better understanding of the viability of this dynamic formation model, and whether it can explain the existence of the eccentric giant population.

\subsubsection{Neptune Desert Planets}
The paucity of planets with sizes similar to that of Neptune and orbital periods below $\sim$4 days, lead to the term Neptune 
Desert (or planet desert) being coined \citep[e.g.][]{howard12,mazeh16}.  The shape of the Desert is driven by the populations of larger and more massive hot Jupiters, 
and smaller and less massive super-Earths, and any theoretical model to explain its existence needs to explain both the upper 
and lower boundaries.  Currently, models that invoke photoevaporation of Neptunes to describe the lower boundary, and tidal 
disruption of tidally migrating Jupiters to explain the upper boundary, do a reasonably good job at explaining the Desert's morphology \citep[e.g.][]{owen13,owen17}.  Indeed, observations of planets around the upper perimeter of the Desert may be pointing
towards effects beyond only photoevaporation as being responsible \citep{Vissapragada21}.

The recent discovery of a few interlopers into the Neptune Desert has given rise to new laboratories to better understand, not 
only the physics of exo-Neptunes in extreme environments close to their host stars, but also the existence of the Neptune Desert 
itself.  For example, questions about additional processes that are expected to be at play in these regions can be addressed, 
such as what role does Roche Lobe Overflow \citep{Valsecchi15} play in the formation of the Desert?  The Tianyu project is not only expected 
to find new Neptune Desert planets to increase the statistical population, but can also search for outer massive companions to 
these worlds, which will allow us to test if these planets could, or could not, previously have been gas giants that migrated too 
close to their host stars and lost the bulk, or all, of their gaseous atmospheres.

\subsection{Variable sources}

\begin{figure*}
	\centering
	\includegraphics[width=0.9\textwidth]{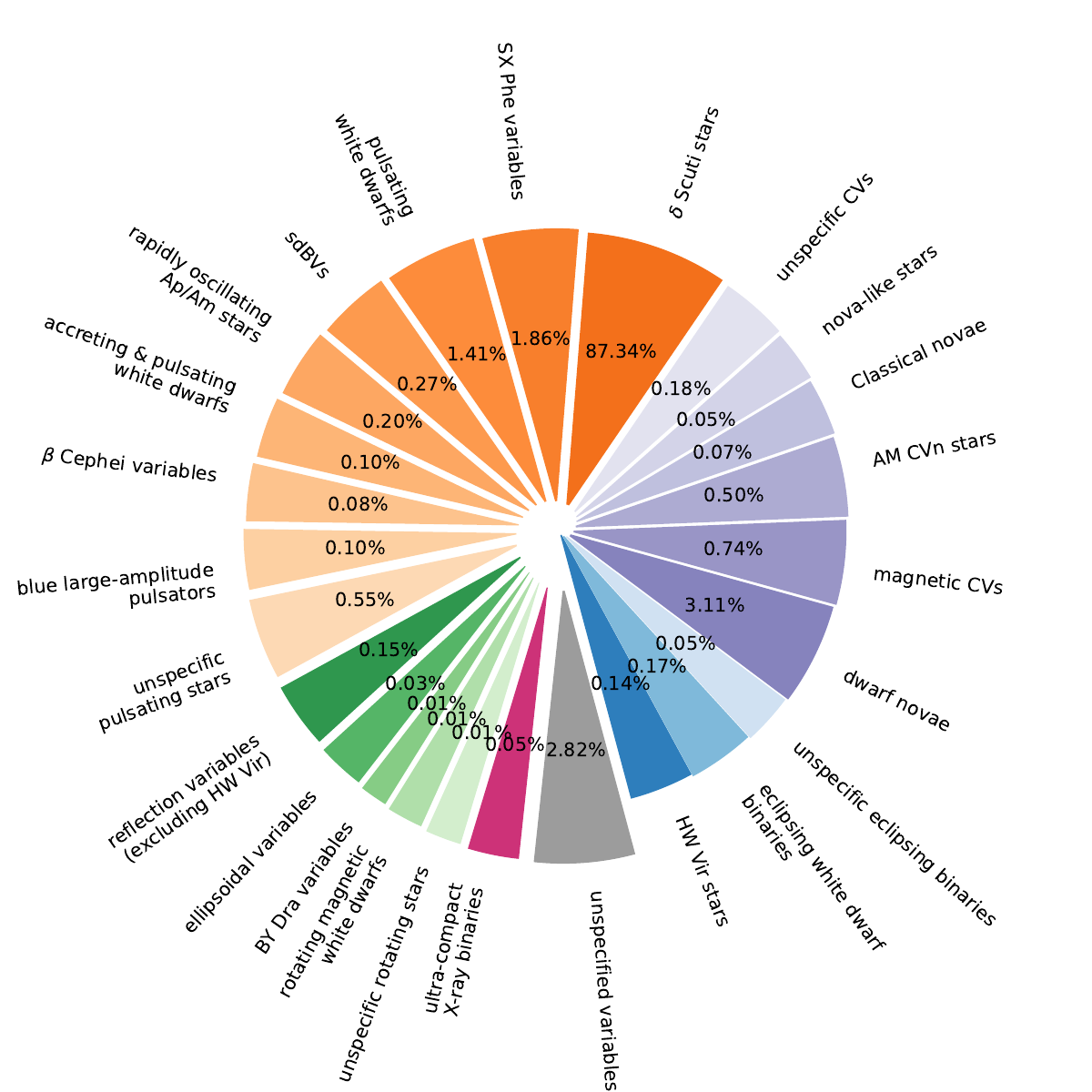}
	\caption{
		Pie chart for fractions of various classes of short-period ($P<2$~hr) variable stars according to the International Variable Star Index (VSX, \cite{Watson+etal+2006+VSX}).
		The pie chart is designed following the Fig.~1 of \cite{Lin+etal+TMTSII} with eclipsing binaries, CVs, pulsating stars, rotating variables, and UCXBs are colour coded with purple, blue, green, orange, and violet red, respectively.
		The version of VSX was updated on December 11, 2023.
		The ``unspecified'' represents those variables lacking further identifications.
	} 
	\label{fig:vsx_statistic}
\end{figure*}

The minute-cadence photometry observations with large-field survey telescopes are very efficient to reveal short-period variable stars among more than one billion observed stars in our Galaxy \citep{Kupfer+etal+2021+ZTFhighcadence,tmtsI+2022}.
As introduced by \cite{Lin+etal+TMTSII}, all short-period variable stars (with period shorter than 2 hours) can be divided into five main types, namely eclipsing binaries, cataclysmic variables, pulsating stars, rotating variables, and ultracompact X-ray binaries (see the Fig.~\ref{fig:vsx_statistic}). These short-period variable stars are important for exploring new pulsation instability strips, studying the extreme evolutionary stages of binaries, and even severing as verification binaries for space-borne gravitational wave (GW) detectors. 

\subsubsection{Pulsating stars}\label{subsec:pulsatingstar}

\begin{itemize}
	\item \textbf{Blue large-amplitude pulsators (BLAPs):} 
	BLAPs are a new and rare class of hot pulsating stars with unusually large amplitudes and short periods, first discovered by Optical Gravitational Lensing Experiment (OGLE, \citep{Pietrukowicz+2017+OGLE_BLAPs}). Their high amplitudes challenge the current asteroseismology theories for hot stars \citep{Corsico+etal+2019+book+pulsatingWD}. Due to the locations of these pulsating stars in the Hertzsprung-Russell diagram (HRD) and Kiel diagram ($T_{\rm eff}-\log\,g$ diagram), the physical origin of BLAPs is thought to be either helium-core pre-WDs or core helium-burning (CHeB) subdwarfs \citep{Wu+etal+2018+CHeB_BLAP,Byrne+etal+2018+BLAPs,Byrne+etal+2020+faint_BLAP}.
	However, the discovery of an 18.9\,min BLAP (TMTS-BLAP-1) with an unusually large rate of period change suggests that the BLAPs are more likely to be shell-helium burning subdwarfs \citep{lin23,Xiong+etal+2022+SHeB}.
	{As one class of the rarest pulsating stars, time-domain survey missions can discover a confirmed BLAP from per 70 million stars \citep{Pietrukowicz+2017+OGLE_BLAPs,McWhirter+Lam+2022+blap_candidates}.
		Hence, by monitoring eight hundred million stars within five years, Tianyu is expected to reveal about 12 BLAPs.
		However, given the high-cadence observation mode, Tianyu has the potential to discover more BLAPs. 
	}

	\item \textbf{High-amplitude $\delta$ Scuti star (HADS):} 
	HADS are a subclass of $\delta$ Scuti star and considered as the transition objects between the main sequence $\delta$ Sct stars and the more-evolved classical (i.e. radial) pulsators, providing great opportunity to investigate the stellar internal physics, such as rotation, overshooting, interaction between binary and pulsations et al. \citep{2009pfer.book.....M,2010aste.book.....A,2021RvMP...93a5001A} . Recently, observational studies from Kepler mission \citep{2010Sci...327..977B} show that HADS possess diverse pulsating behavior, such as amplitude modulation and decline, which may originate in stellar rotation and pulsation energy loss, challenging our understanding of pulsation theories \citep{2018ApJ...863..195Y,2022ApJ...936...48Y,2023ApJ...955...80S}. New observations also reveal that HADS can also be in binary system undergone mass transfer, providing an opportunity for testing the current theories of binary evolution, mass transfer, and pulsation \citep{2021A&A...655A..63Y}. The stellar companion orbiting $\delta$ Sct variables can be identified using the phase modulation technique \citep{2022Univ....8..614Y,2023ApJ...943L...7L}. {By monitoring more than 200,000 stars in about 115 deg$^2$, Kepler mission has found 5 HADS \citep{2016MNRAS.460.1970B,2019ApJ...879...59Y,2021AJ....161...27Y,2022ApJ...936...48Y}. Since Tianyu is planned to monitor eight hundred million stars within five years, we expect to discover about 20 thousands HADS with the observations of Tianyu. However, given the high-cadence observational strategy and larger sky-zone, Tianyu has the potential to find more HADS and thus provide more clues to understand mode selection, stellar rotation and pulsational diversity.}

\end{itemize}

\subsubsection{Eclipsing/ellipsoidal binaries}\label{subsec:EB}

\begin{itemize}
	\item  \textbf{Verification binaries for GWs:}
	With the operations of space-borne GW detectors \citep{LISA+2017,Luo+etal+2016+TianQin} in the future, a large number of ultracompact binaries (UCBs) can be detected through GW messengers within their first-three-month observations. The UCBs that have an orbital period shorter than tens of minutes will server as verification binaries for space GW observatories \citep{Kupfer+etal+2018+VBs,Kupfer+etal+2023+LISA}.
	
	By monitoring more than one billion stars in the Milky Way, Zwicky Transient Facility (ZTF) has discovered 50 UCBs, among which four UCBs are the verification binary of gravitational waves \citep{Burdge+etal+2019+7min,Burdge+etal+2020+systematic_VBs,Kupfer+etal+2023+LISA}.
	Since Tianyu is planned to monitor eight hundred million stars within five years, we expect to discover about 40 UCBs with the observations of Tianyu.
	However, Tsinghua University-Ma Huateng Telescope for Survey (TMTS) \citep{Zhang_2020,tmtsI+2022} revealed a verification binary from only ten million light curves based on minute-cadence observations \citep{Lin+etal+TMTSII,Lin+etal+20.5min},  which has a very high efficiency in searching for short-period variable stars.
	Given the short-cadence observation mode of Tianyu, the number of UCBs here could be severely underestimated.

	
	\item  \textbf{Binaries during extreme/transitional evolutionary stages:} 
	The binaries in extreme/transitional evolutionary stages can provide key evidences for current binary evolution scheme \citep{Han2020RAA....20..161H}, e.g. the hydrogen-star channel of AM CVn stars \citep{Nelemans+etal+2001+AMCVn,Solheim+2010+AMCVn,Burdge+etal+2022+TCV+nature}.
	Since such binary systems are very rare and usually short-orbital-period, searching for these binaries requires the telescopes with large FOV and high observation cadence.
	As a telescope with large FOV and multi observation cadences, Tianyu is expected to play a key role in searching and studying special binaries predicted by binary evolution theories in the future.

	\item  \textbf{Ultracompact black hole binaries (UCBHBs):} 
	As the ultracompact binaries harbouring a black hole (BH) and a visible star (e.g. white dwarf or sdB star), UCBHBs are expected to be detected by both electromagnetic and gravitational waves \citep{Chen+etal+2020+UCXB_LISA,Wang+etal+2021+UCXB}, thus providing the most direct evidence that stellar-mass black holes exist.
	However, among tens of known ultracompact X-ray binaries (UCXBs), only 47~Tuc~X9 \citep{Bahramian+etal+2017+BH_UCXB} was suspected to harbour a BH accretor while others were identified as neutron star (NS) systems.
	Since the missing of BH UCXBs are inferred to be caused by radiatively inefficient accretion \citep{Knevitt+etal+2014+inefficient_accretion} or very long recurrence time of X-ray outbursts \citep{Lin+etal+2019+outbursts}, these short-orbital-period noninteracting (or quiescent) BHBs are likely discovered by periodic ellipsoidal variations in light curves from optical photometry surveys by Tianyu. While the frequency of UCBHBs remains inadequately researched, Tianyu's extensive survey encompassing tens of millions of stars is poised to uncover further instances of these enigmatic phenomena. 
	
	\item  \textbf{Solving eclipsing binaries with a solar-like pulsator:} 
	Binaries play a crucial role in examining and refining stellar evolutionary theories. Numerous eclipsing binary stars have been discovered through ground-based projects, including those from the Antarctica \citep{2015ApJS..217...28Y,2019ApJS..240...16Z}. Eclipsing characteristics offer valuable orbital information, along with the relative brightness and radii of the two component stars. When one component star exhibits detectable solar-like oscillations, the entire system becomes solvable. Scaling relations based on solar-like oscillations have been proven effective in estimating stellar mass and radius \citep{2011ApJ...743..143H,2013ApJ...767..127H}. Therefore, a more comprehensive understanding of eclipsing binaries with a solar-like pulsator can be achieved through a combination of asteroseismic analysis and binary modelling \citep{2013ApJ...767...82G,2014ApJ...785....5G,2019MNRAS.487.2455O}. Observing solar-like oscillations from the ground is challenging due to their low amplitude and high pulsation frequency. However, employing a 1-minute short-cadence greatly aids in resolving high-frequency pulsations. Red giants, in particular, exhibit relatively larger pulsation amplitudes compared to main-sequence stars. This makes Tianyu capable of observing red giant pulsations. If the red giant happens to be part of an eclipsing binary system, we can obtain precise orbital and physical parameters for the entire system by combining asteroseismology and eclipsing geometry. Additionally, the stellar masses can also be determined through radial velocity measurements. These independent measurements of stellar parameters make such systems valuable for testing stellar evolution models. While the All-Sky Automated Survey for Supernovae (ASAS-SN) has identified 649 red giants in eclipsing binaries \citep{rowan22,rowan23}, Tianyu's surveying capacity, spanning eight times more stars than ASAS-SN, is anticipated to uncover thousands more.
	
\end{itemize}

\subsubsection{Active stars}
\begin{itemize}
	\item \textbf{Rotators:} Rotation modulation, commonly observed in time-domain surveys, is considered a prominent feature in light curves. Both ground based missions and space telescopes have identified numerous rotators with period {$P_\mathrm{rot}$} varies from ${\sim}0.1$ to 100 days \citep[e.g.,][]{mcquillan_rotation_2014, newton_new_2018, cui_long_2019}. Their rotation periods and activity indicators provide valuable information about stellar age, magnetic field, and the potential habitability of associated planets. On one hand, rotational modulation, as the primary source of intrinsic noise, identifying and subtracting them is crucial for the detection of transit signals. On the other hand, independent studies on stellar activity also hold significant value. {Tianyu project will provide a greater number of rapidly rotating stars ($P_\mathrm{rot}<10$\,days) in its short and medium-cadence observation modes. By combining previous multi-band observations and spectra survey like LAMOST \citep{Cui:2012:RAA}, we can examine the evolution of their activity. Furthermore, our long-cadence mode will complement TESS's short observational seasons by identifying numerous slow rotators ($P_\mathrm{rot}> 30$\,days). Their expected lower activity level makes them suitable hosts for detecting even weaker transiting signals.}
	
	\item \textbf{Flares:} Flares are significant increases in brightness that are observed on light curves, also attributed to magnetic activities. High-precision space photometric observations from Kepler and TESS have enabled the detection of white-light flares, including those with {notably short durations}, which were often missed in previous space missions \citep{2020ApJ...890...46T,2021ApJS..253...35T,aizawa_fast_2022,2022AJ....164..213G}. 
	{ Our Tianyu project specifically leverages an ultrashort exposure time capability, distinguishing itself by detecting flares on a ultra short timescales---down to ${\sim}0.3$\,s. This not only significantly increases the volume of detectable ultra-fast flares (${\sim}1$\,s) but also offers a unique insight into their intricate mechanisms, which could be crucial for understanding stellar magnetic activities and their potential impacts on exoplanetary environments.
	}

	
\end{itemize}

\subsubsection{ Be stars}

Be stars are fast-rotating early-type stars which show or once shown emission in their Balmer lines \citep{Collins1987}.
Such emission stems from the circumstellar disk formed during the Be phenomena, where the material in the equator of the stars is lifted up and forms the so-called decretion disk.
The optical brightness of the Be star may increase or decrease when the decretion disk forms, depending on the viewing angle between the line-of-sight and the stellar rotation axis \citep{Haubois2012}.
Recently, \cite{jian2023} discovered more than 900 Be stars with Be phenomena in the Milky way using the data from the Wide-field Infrared Survey Explorer (WISE; \cite{Wright2010}).
They will serve as the main targets and being observed by the long-cadence survey of Tianyu.
Such dataset provides photometry observation with a much shorter cadence compared to the one from \cite{jian2023}, allowing us to capture the short-term brightness variation which will be missed from the WISE data.
The brightness variation in the Tianyu band (i.e., optical) is more sensitive to the inclination angle than the WISE bands, thus they can be used to constrain the inclination angle of the target Be stars, which can hardly be measured without the optical light curve.
The wide FOV of Tianyu is capable to observe the already-found Be stars and detect new ones with high efficiency.
Based on the number of newly detected Be stars in \cite{jian2023} and Tianyu's observation strategy, we estimate that we can measure the rotation axis angle of around 250 stars and detect around 100 new Be stars after its first 5-year survey. 

\subsubsection{Active Galactic Nucleus}
\begin{itemize}
	\item\textbf{BH seeds:} Variability is ubiquitous in Active Galactic Nuclei (AGNs) and serves as a unique feature for their identification \citep{Ulrich97}. As observations push the discovery of first-generation quasars to earlier and earlier cosmic epochs, how supermassive black holes (SMBH, $M_{\rm BH} \ge 10^6 M_{\odot}$) formed becomes a very urgent problem in cosmology that needs to be resolved. At least three channels have been proposed for the formation of the seeds of SMBHs: Pop III stellar remnants \citep[e.g.,][]{Madau01}, direct collapse \citep[e.g.,][]{Bromm03}, or star cluster evolution \citep{Miller02,PortegiesZwart02}. Probing BH seeds in the early universe, however, is beyond any current observational capability, while fast growing SMBHs at high-z quickly loose all record of their origin \citep{Valiante18}. Fortunately, theoretical work suggests that observations of local leftover Intermediate-Mass BHs (IMBHs, $10^2 - 10^6 M_{\odot}$) in dwarf galaxies (i.e., stellar mass $M_{\star} < 10^{10}M_{\odot}$) may also provide some crucial clues to understand seeding processes, since they suffered relatively fewer mergers and less accretion since their formation than their supermassive peers \citep{Volonteri08}. With the operations of Tianyu, there is potential to uncover new IMBHs in nearby dwarf galaxies, benefiting from its short-cadence monitoring and wide FOV.  
	
	\item\textbf{Continuum reverberation mapping.} Constrained by the spatial resolution limitations of current facilities, the accretion disks around massive black holes (BHs) are generally too small to be directly resolved, even for the closest AGNs. The technique of reverberation mapping (RM) overcomes this constraint and emerges as a primary tool for unveiling the physics and environment surrounding massive BHs \citep{Blandford82}. Within the framework of the X-ray reprocessing model \citep{Krolik91, Cackett07}, continuum reverberation mapping (CRM) aims to identify delays among different continuum bands, effectively mapping the physical distances between various annuli of accretion disks. This technique enables the measurements of accretion disk sizes and facilitates the study of mechanisms contributing to X-ray to infrared variability \citep{Cackett21}. The scientific goal is to constrain the disk size and verify the capability of the standard accretion disk model and the X-ray reprocessing model in AGNs. Given the relatively small aperture size of Tianyu, it is well-suited for coordinated monitoring of a single broad band in CRM campaigns and measuring optical continuum inter-band lags. Specifically, Tianyu-II will observe about 100 AGNs at $z<1$ with a cadence of a few days to measure CRM with broad band filters.
\end{itemize}

\subsection{Transients}

With the new generation of all-sky optical surveys, a large number of transients have been detected and a fraction of unexpected transients challenge the existing stellar evolution models. Tianyu project, including the Tianyu-I and Tianyu-II telescopes, is an essential observational tool for testing stellar evolution models.  

\subsubsection{Supernovae}\label{subsec:SN}
The current research focus in supernova research lies in the earliest observations, i.e., infant supernovae, ranging from a few hours to a day post-explosion. Data from this period maximally retains information from just before and at the moment of the supernova explosion, shedding light on the properties of progenitor stars and the physical mechanisms of the explosion \citep[e.g.][]{2018Natur.554..497B,2023arXiv231114409L}. Given the unpredictability of both the timing and location of supernova explosions, and the concentration of early-time radiation in the ultraviolet to optical band, the ideal observational strategy is through wide-field surveys in corresponding wavelengths. The survey strategy of the Tianyu project is well-suited to meet this requirement.

The medium-cadence survey program can timely discover supernovae within a few hours after the explosion. With real-time data processing and transient source alerts, early multi-band photometry and/or spectroscopy can be conducted using the Tianyu-II telescope. Early multi-band photometry of core-collapse supernovae can be used to analyze whether there is circumstellar material around the supernova, thereby providing clues for the study of the late evolutionary stages of stars \citep[e.g.][]{2023arXiv231114409L}. Early observations of Type Ia supernovae can 
constrain the types of progenitor stars and reveal the explosion mechanisms
such as delayed detonation and double detonation \citep[e.g.][]{2023MNRAS.525..246H,2023MNRAS.522.6035M,2021ApJ...906...99L,2020ApJ...902...48B,2017Natur.550...80J}.

\subsubsection{Intermediate Luminosity Optical Transients (ILOTs)} 
ILOTs are one of the new astrophysical frontiers, whose intrinsic luminosities are fainter than normal supernovae (lower limit magnitude $\sim$ $-$15) but brighter than typical novae (upper limit magnitude $\sim$ $-$10). For this reason, these transients are often called `gap transients' \citep[e.g.][]{Kasliwal2012PASA...29..482K, Pastorello2019NatAs...3..676P,Cai2022Univ....8..493C}. This luminosity gap is populated by several categories of stellar transients, such as different types of under-luminous supernovae \citep[e.g.][]{Pastorello2004MNRAS.347...74P, Valenti2009Natur.459..674V, Spiro2014MNRAS.439.2873S}, giant eruptions/outbursts of massive stars including Luminous Blue Variables \citep[LBV; e.g.][]{Humphreys1994PASP..106.1025H, Smith2011MNRAS.415..773S,Humphreys2016ApJ...825...64H}, intermediate luminosity red transients \citep[ILRTs; e.g.][]{Botticella2009MNRAS.398.1041B, Cai2018MNRAS.480.3424C, Cai2021A&A...654A.157C}, and luminous red novae \citep[LRNe; e.g.][]{Pastorello2019A&A...630A..75P, Cai2019A&A...632L...6C, Han2020RAA....20..161H, Cai2022A&A...667A...4C, Ge2023arXiv231117304G, Chen2024PrPNP.13404083C}. They show heterogeneous observational properties, formed in various progenitor channels, and triggered by different explosion mechanisms. Although modern all-sky surveys discovered some ILOTs in past few years, there is only a limited amount of events have been observed with well-sampled and high-quality data sets, hence our knowledge on the nature of ILOTs is still incomplete. 

In the context of the Tianyu project, we expect significant improvements on the ILOTs science: 1) Detecting the pre-outburst light-curve variability. Through stacking multiple exposures, we can go much deeper in magnitude that allows us to inspect the progenitors of ILOTs (e.g. eruption phase in LBVs; periodic variability in LRNe). 2) Discovering new ILOTs events. The increasing number of ILOTs can help us determine their precise occurrence rates and constrain their progenitor scenarios. For example, ILRTs rate ($\sim$ 8\%) is consistent with the electron-capture (EC) supernova theoretical rates ($\sim$ 2-10\%), suggesting their EC supernova nature \citep[see details in ][]{Cai2021A&A...654A.157C}. 3) Monitoring the ILOTs with routine observations. Well-sampled light curves and colour evolution will help us classify, characterise, and analyze the ILOTs properties. Therefore, Tianyu (I\&II) data will play a major role in studying the metallicity, binarity, progenitor, and mass-loss history of ILOTs. 

\subsubsection{Kilonovae}
A new era in multi-messenger astronomy was launched by the first detection of both gravitational
waves and electromagnetic radiation from the same event - the neutron-star merger GW170817 accompanied by a kilonova AT\,2017gfo
\citep{Abbott2017c_170817gws,Abbott2017b_170817mma,Arcavi2017c_kn}. This dual detection is providing new ways to probe physics from the sub-atomic scale to cosmology \citep{Abbott2017a_hubble}.

Optical-infrared emission from a neutron-star merger was predicted to originate from the radioactive decay of heavy elements produced from a small amount ($10^{-4}$--$10^{-2}M_\odot$) of ejecta \citep[e.g.][]{Li1998,Rosswog2005,Metzger2010,Roberts2011}. The sources of the ejecta, which could be from the tidal tails, the contact interface of the two neutron stars \citep[e.g.][]{Hotokezaka2013}, or winds from a post-merger accretion disk \citep[e.g.][]{Metzger2008,Grossman2014,Metzger2018}, play a crucial role in determining its mass, velocity and composition. These factors, in turn, are reflected in the characteristics of the emission properties \citep[e.g.][]{Kasen2013,Tanaka2013,Barnes2013,Grossman2014}. 

The gravitational wave detectors LIGO, Virgo, and KAGRA will conduct the O5 Observing run during 2027-2029, which coincides with the operational period of Tianyu. Both LIGO and Virgo will be able to observe neutron-star merger events within 150\,Mpc, enhancing the precision of localisation. During the O5 campaign, it is expected that dozens of neutron-star merger events will be discovered. For events with better localisation precision, the accuracy can reach tens of square degrees. Alerts with sky maps will be issued in real-time. Tianyu-I can scan the sky area covered by the gravitational wave signals in ToO mode, covering the localisation area within a few exposures. If rapid image processing and analysis can be performed, Tianyu-II can conduct multi-band photometric follow-up observations of potential kilonova candidates, enabling us to study the energy source problems mentioned above through photometric analysis.

\subsubsection{Tidal disruption events}\label{subsec:TDE}
When a star passes by a massive black hole (BH) close enough, the tidal force from the BH exerts on it can be larger than the self-gravity of the star. It leads to the tidal disruption event of the stars (TDE). The dynamics of the residual debris is closely associated with the impact parameters, as well as the size and mass of both the BH and the star. The gravitationally bound debris accretes into the black hole at an exceptionally high rate, surpassing the Eddington limit by a considerable margin. Rapid accretion generates highly non-linear process and a highly variable light curve \citep{Sadowski2016, 2023MNRAS.521..389R,Andalman2022,2023MNRAS.522.2307J}. Rapid response and continued observation will be important to the study of TDE. A large portion of TDEs have been discovered via wide-field optical surveys in the past decade \citep{2021ARA&A..59...21G}, which makes the wide FOV of Tianyu-I rather important in discovering TDE.

\subsubsection{Fast blue optical transients}
Recent short-cadence surveys have discovered a new kind of rapidly evolving optical transients with diverse observed properties compared to typical supernovae. 
The short evolution timescales, high peak luminosities, and persistent blue colors as seen in those optical transients are in
tension with traditional supernovae models powered by the radioactive decay of 56Ni. In particular, the short rising timescales imply a low ejecta mass ($M_{\rm ej}\ll 1M_{\odot}$) with a high velocity ($v_{\rm ej}\gg 0.05c$). However, the high luminosity peak requires plenty of 56Ni that could be much larger than that of ejecta mass, which suggests diverse origins of energy sources.
Those fast blue optical transients (FBOTs) challenge our knowledge on stellar evolution and their death \citep[e.g.][]{drout14,rest18}. 

\subsection{Solar system objects}

The focus of Tianyu for solar system observations is the small bodies, including asteroids and comets. Small bodies are the remnants of the assembly of planets, and as such, record the chemical composition and physical conditions of the planetary formation process, as well as the signatures of the dynamic migrations of giant planets in the early solar system (e.g., \citep{gomes05,2011Natur.475..206W}).

Broadly speaking, the solar system's small body population includes NEOs, Main-belt asteroids (MBAs), Jovian Trojans, Centaurs, and TNOs.  NEOs are those with obits crossing or close to the Earth's orbit, including both asteroid and comet populations. They are the major sources of impactors into the Earth's atmosphere and may have transported volatiles and organic matter to shape the habitable environment on early Earth \citep{1961Natur.190..389O}. NEOs could potentially cause catastrophic impacts on Earth (e.g., \citep{2022NRvEE...3..338M}) and are therefore the main subject of planetary defence. They are also potential space-born resources for future space explorations. MBAs are distributed between Mars and Jupiter and are considered the source of near-Earth asteroids through various dynamic resonances \citep{2002aste.book..379N}. Jovian Trojans are a dynamically stable population distributed in the L4 and L5 Sun-Jupiter Lagrange points with a lifetime exceeding the age of the solar system but an unknown origin \citep{2003Icar..162..453M}. Centaurs are distributed in the giant planet region between Jupiter and Neptune and are dynamically unstable with a lifetime of 10$^6$ years. They are generally considered the transitional phase of TNOs towards Jupiter family comets \citep{2009AJ....137.4296J}. TNOs are distributed outside of Neptune and are analogous to the debris disk of planetary formation \citep{2020tnss.book....1F}. Given the 1-m aperture size of Tianyu, its main solar system targets will be NEOs, MBAs, Trojans of $>$10s of km, and the few largest TNOs (assuming a V-band limiting magnitude of ~19, to be checked).

Comets are volatile-rich objects that eject gas and dust in the inner solar system driven by volatile sublimation. Two main dynamic populations of comets exist and are considered to have formed in distinct regions in the solar system and undergone different dynamic paths since formation \citep{2022arXiv220600010K}. The Ecliptic Comets (ECs) have their orbits close to the ecliptic plane and are considered to originate from the trans-Neptunian region through the Centaur phase \citep{fraser2022transition}. The Jupiter family comets (JFCs) are part of the EC population and have their aphelia near Jupiter's orbit. The nearly isotropic comets (NICs), on the other hand, have their orbits almost isotropically distributed between 0$^\circ$ and 180$^\circ$, and are likely originated from the hypothetical Oort cloud at $>$1000 au from the Sun. Most NICs have orbital period $>$100s to 1000s of years, namely long-period comets. Some of the NICs that originate from $>$200,000 au from the Sun likely enter the inner solar system for the first time after their formation and are therefore called dynamically new comets. They are perhaps the most pristine samples from the solar system formation and carry unique scientific information. Due to the small nucleus size of typically a few km \citep{2023arXiv230409309K}, comets are mostly discovered after they have developed cometary activity and showed the fuzzy appearance of the coma and tail. Comets display extremely diverse volatile and refractory compositions and the connection to their origin is still poorly understood (e.g., \citep{2011ARA&A..49..471M,2022arXiv220704800B}). However, the opportunity to observe particular comets, especially long-period comets, is very limited. Therefore, it is critical to discover as many comets as possible and closely monitor their behaviors throughout their inner solar system passages in order to increase the sample size and study the diversity of comets.

Tianyu will primarily contribute to the discovery of previously unknown small bodies, especially NEOs and comets, and the photometric characterization of known asteroids and comets. Considering 231 NEOs and comets discovered by ZTF to date, we estimate that the number of NEOs and Comets can be discovered by Tianyu in six years is 58, as shown in Table \ref{tab:yield_other}.

\subsubsection{NEOs, MBAs, and TNOs}\label{subsec:NEO}
With its large FOV and varying observation cadences, Tianyu will capture the motion of small bodies with a vast range of apparent speeds, hence contribute to the discovery of previously unknown objects as well as further characterisation of those known. Automated algorithm will be developed to examine difference images for the extraction of both streaked and point-like moving targets, from which light curve data will be assembled. The accumulated astrometric and photometric data will be applied for the precise orbit determination as well as the characterisation of physical properties such as size, shape, rotation and composition. 

The rotation and shape of asteroids are important parameters in the study of solar system objects \citep{2002aste.book..113P}. Understanding the rotational axis orientation and shape of an asteroid provides valuable insights into its internal structure and the distribution of its constituent materials. Despite advancements in observational technologies, accurately determining the rotation axis and shape of asteroids remains a challenge. This is due to the requirement for continuous time-domain observation data, ensuring an adequate time span, sufficient data, and an appropriate model to distinguish and resolve ambiguities \citep{2001Icar..153...24K,2001Icar..153...37K}. Tianyu is expected to contribute significant data to the field of asteroid observation. With its capabilities, Tianyu can conduct detailed analyses of specific targets, including obtaining information on rotation axis direction and shape. This will greatly support the strategic needs of China's deep space exploration initiatives.

Potential investigations include:

\begin{itemize}
	\item \textbf{PHA:} Potentially hazardous asteroids (PHAs) are NEOs with orbits that can approach to the Earth within 0.05\,au. Depending on the size, their impacts on the Earth could cause regional or global damages that poise existential crisis to life. Early discovery and precise characterisation of PHAs are key for mitigating their threat. 20\% of NEOs are PHAs that pass within 0.05 au of the Earth. Less than 50\% of NEOs smaller than 140\,m have been discovered. 
	
	\item \textbf{Asteroid taxonomy:} The spectra of asteroids are determined by their surface mineralogical composition. Based on the characteristics of the asteroid spectra, they are classified into taxonomic systems, the most recent and widely adopted being the Bus-DeMeo system \citep{2009Icar..202..160D}. Multi-filter photometry of asteroids will provide information about the taxonomy classes of asteroid populations. Tianyu is well-positioned to provide a large database of asteroid colors to support the study of asteroid surface composition, space weathering, and asteroid families. The color information of newly discovered asteroids will provide the first constraints on their possible compositions and estimate of albedo, and therefore better constrain the size from their brightness.
	
	\item \textbf{Yarkovsky and YORP effect:} Orbits and rotation states of small bodies are subject to long-term variations caused by the so-called Yarkovsky and YORP effect, respectively \citep{2006AREPS..34..157B}. Though subtle, these thermal forces are crucial to an asteroid's dynamic evolution on the time scale of millions of years. They could perform as the decisive factor for the transport of NEOs from the main belt. Due to still limited measurements, our current knowledge of these effects are largely dependent on theoretical models. Tianyu's data will expand the sample size of Yarkovsky and YORP measurements, improving our knowledge of their role in shaping the distribution of small bodies.
	
	\item \textbf{Binary system:} A binary system, such as the Didymos system targeted by NASA's DART mission \citep{2023Natur.616..443D}, is a pair of asteroids orbiting around their common center of mass. About 15\% of NEAs larger than 300\,m are binaries \citep{2006Icar..181...63P}, and the fraction of binaries among MBAs is not well established, but is suspected. They could be of diverse origin, including collisions, rotational breakup, and tidal disruption. Photometric light curve decomposition has been widely used as an effective method to characterise binary systems (e.g., 
	\citep{2012Icar..218..125P}). The database of small body light curves by Tianyu will help identify and constrain the configuration of binary systems, for assessing different formation theories and revealing their unique dynamic paths through the history of the solar system. 
	
	\item \textbf{TNO:} With a limiting magnitude of 21, Tianyu is able to only directly observe the largest TNOs. However, there are chances that it could provide serendipitous occultation observations that occur when a TNO passes in front of a star and dims its light (e.g., \citep{2021PASP..133c4503H}). The resulting transit light curve could be applied in the estimate of the TNO's size as well as searching for possible ring structures around it. Furthermore, coordinated TNO occultation campaigns with multiple telescopes could reveal the cross-section of the target for the constraint of its shape. 
	
	\item \textbf{Space mission targets:} The data inventory of small bodies built by Tianyu will also help with the selection and characterization of targets for future space missions. Small body missions could be of diverse goals including scientific exploration, technological demonstration for planetary defence, space resource utilisation, and so on. The successful evaluation and selection of targets according to the mission intent depend on comprehensive database on the knowledge of candidate targets. Further detailed information on the selected target is needed during the development phase of a mission to guarantee an optimised design of mission profile as well as hardware payloads.
	
\end{itemize}






\subsubsection{Active small bodies}

Comets are classical active small bodies characterised by mass loss in both gas and dust. The driving mechanism for cometary activity varies with heliocentric distance. Within 3\,au, the mechanism of comet activity is overall consistent with the standard model \citep{1950ApJ...111..375W}, and water ice sublimation is the main driving force. Beyond 5 au, the sublimation of ``super volatile'' ices such as CO$_2$ and CO could become dominant \citep{1993PASP..105..946L}, and different driving modes drive the release of dust from the surface of the comet nucleus to form different activity of the comet. Therefore, the study of comet activity at different heliocentric distances is expected to shed light on the driving mechanism of comet activity.

Active asteroids are relatively new additions to active small bodies. They were fist discovered in the main asteroid belt, where some asteroids were found to exhibit comet-like coma and\/or tails \citep{2006Sci...312..561H}. A range of mechanisms could be responsible for driving the activities, such as the sublimation of water ice \citep{2023Natur.619..720K}, impacts \citep{2012AJ....143...66J,2023Natur.616..452L}, rotational destabilisation \citep{2015ApJ...798..109J}, radiation pressure sweeping, electrostatic ejection, thermal fracture, interplanetary space interaction, etc. The existence of sublimation-driven activity in the main asteroid belt revealed the new inventory of water ice in the solar system between Mars and Jupiter and provided the approach to study the volatile distribution in the inner solar system and transportation into the terrestrial planets. The other activation mechanisms in the asteroid belt opened a new area for studying the interaction between asteroids and the production of interplanetary dust in the solar system and the formation and evolution of the zodiacal cloud.

The activity of active asteroids is weak \citep{2018AJ....156..223H,2019NatSR...9.5492S}, the number of discovered active asteroids is 37 \citep{2024MNRAS.52710309X}, of which the number of active asteroids driven by water ice sublimation is 15 \citep{2023Icar..40115605L,2016ApJ...830...22H,2021ApJ...922L...9H}. There may be great uncertainty in extrapolating the overall character of active asteroids by analysing the orbits and activity characteristics of discovered active asteroids. Therefore, the current research focus is to expand the sample of active asteroids and formulate reasonable observation and search schemes for active asteroids. Many survey telescopes are involved in the search for active asteroids, such as Pan-STARRS1 (PS1; \cite{Chambers2016}), the Hawaii Trails Project (HTP; \cite{Hsieh09}), the Canada  France-Hawaii Telescope (CFHT; \cite{Gwyn_2012}) and the Palomar Transient Factory (PTF; \cite{2009PASP..121.1395L}). 

Tianyu-I has a 10-square-degree FOV and both its survey strategy and survey time mode are suitable for studying the long-term evolution of comet activity as well as searching for active asteroids. With Tianyu-II, we can conduct detailed multi-band photometry and morphology of specific comet targets to study their physical properties and characterise their activity. In addition to broadband filters, Tianyu-II can also utilise narrow-band filters to better isolate dust and specific gas species. These filters can be applied to better constrain the production rates of dust and gas, hence better measure the dust-to-gas ratio, which is a fundamental compositional parameter of comets related to their origin. Furthermore, the morphological features in gas comae usually have larger spatial scales than those in dust comae due to the higher velocity of gas species. Because those features are usually directly modulated by the rotation of the nucleus and the production of particular gas species from the nucleus or dust grains in the coma, they can provide clues about the fundamental properties of the nucleus and the sources of gas species (e.g. \citep{2004come.book..449S}). The most widely adopted narrow band filters at present are the HB filters \citep{2000Icar..147..180F}, the important ones being listed in Table \ref{tab:filters}.

\begin{table}[ht]
	\caption{Some HB Comet filters (from \citep{2000Icar..147..180F})}
	\label{tab:filters}
	\begin{center}       
		\begin{tabular}{lcc} 
			\hline
			\rule[-1ex]{0pt}{3.5ex}  Species  & Central wavelength(nm) & Transmission bandwidth(\r{A}) \\
			\hline
			\rule[-1ex]{0pt}{3.5ex}   OH (0-0) &	309.7 & 58  \\
			\rule[-1ex]{0pt}{3.5ex}   CN ($\Delta \nu=0$) &	386.9 & 56  \\
			\rule[-1ex]{0pt}{3.5ex}   Blue continuum &	445.3 & 61  \\
			\rule[-1ex]{0pt}{3.5ex}   C$_{2}$ ($\Delta \nu=0$) &	513.5 & 119    \\
			\rule[-1ex]{0pt}{3.5ex}   Green continuum &	525.9 & 56  \\
			\hline
		\end{tabular}
	\end{center}
\end{table} 

\subsubsection{Interstellar objects}
The discovery of the first interstellar object `Oumuamua by Pan-STARRS1 open a window for us to understand extra-solar planetary systems. `Oumuamua's unusual characteristics, such as its small velocity relative to the local standard of rest \citep{mamajek17}, its tumbling rotation \citep{drahus18}, and unusual non-gravitational acceleration \citep{micheli18}, sparked immense interest within the scientific community. Unlike `Oumuamua, the second interstellar object 2I/Borisov exhibited clear cometary activity, releasing gases as it approached the Sun \citep{guzik20}. Due to the limited number of detected interstellar objects, the abundance and formation mechanisms of interstellar objects are quite uncertain \citep{feng18c,zwart18,price21}.  

Considering that detectable interstellar objects tend to cluster around the solar apex and ecliptic plane \citep{hoover22}, Tianyu will conduct long-cadence survey (about once per week) in these favoured regions. With a magnitude limit of about $V=22$\,mag, surveys like Tianyu is expected to find at most 1 interstellar object per year according to the simulations conducted for the Legacy Survey of Space and Time (LSST or Rubin observatory; \citep{hoover22}). 

\subsection{Astronomy outreach}
Astronomy has always been a pivotal field for the education of students and the general public in science. The Inter-Commission B2-C1-C2 Working Group on Data-driven Astronomy Education and Public Outreach (DAEPO), under the International Astronomical Union (IAU), was officially launched in April 2017 \citep{cui2018iau}. There have been successful experiences in providing scientific education to international students using robotic telescopes \citep{Coward2017}.

Tianyu-I and Tianyu-II are set to generate a vast amount of data. By leveraging the robust public science education platform of the Shanghai Astronomy Museum, a branch of the Shanghai Science and Technology Museum, extensive public outreach in astronomy can be effectively conducted. This includes guiding the public in the verification of observational data and opening up certain observation times for the general public and astronomy enthusiasts, facilitating the search for asteroids and supernovae. Despite the remoteness of the observation site, the expertise developed from previous robotic telescopes, in tandem with advancements in astronomical artificial intelligence, as exemplified by StarWhisper\citep{StarWhisper}, paves the way for enabling public participation in astronomical endeavors through automated and intelligent approaches.

\section{Observation site, dome, telescope, and instruments}\label{sec:instr}

\begin{figure}
	\centering
	\includegraphics[width = 0.7\linewidth]{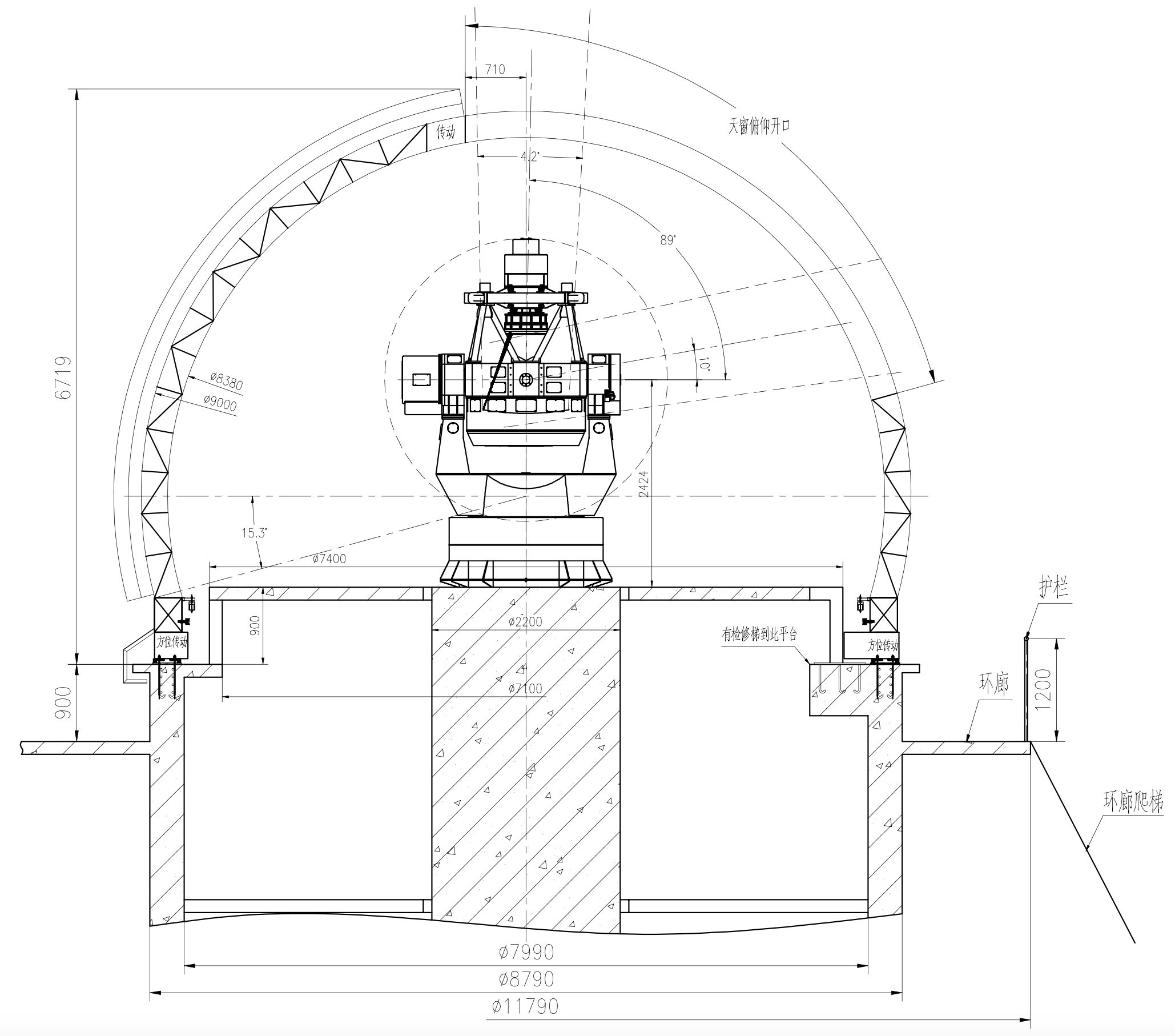}
	\caption{Conceptual design of dome for the Tianyu telescopes.}
	\label{fig:dome}
\end{figure}

\begin{figure}
	\centering
	\includegraphics[width = 0.9\linewidth]{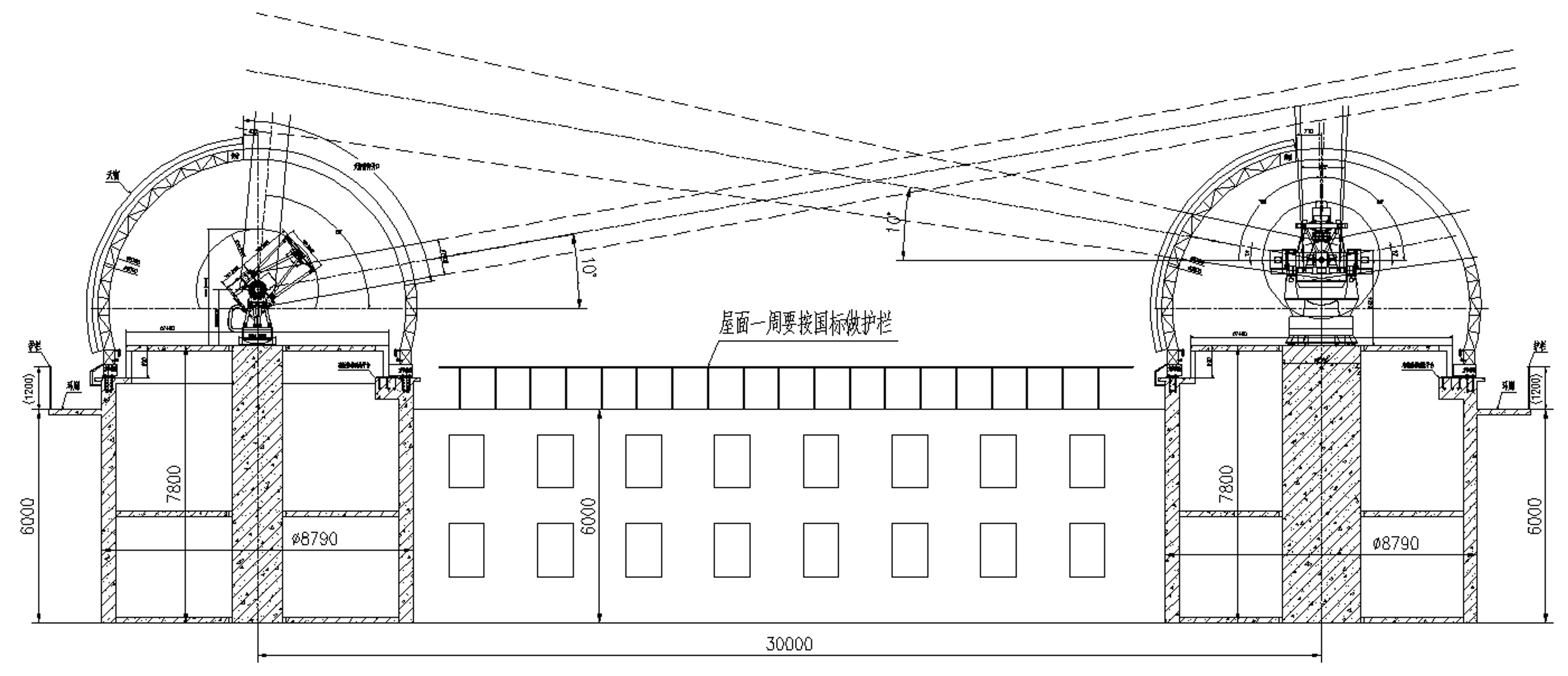}
	\caption{Conceptual design of the infrastructure for the two Tianyu telescopes.}
	\label{fig:infrastructure}
\end{figure}

\subsection{Observation site and dome}\label{subsec:site}

Tianyu will be situated at an altitude of approximately 4000\,m on Saishiteng Mountain in Lenghu, Qinghai, China. Renowned for its extremely arid climate and clear skies, Lenghu offers favourable observing conditions. As reported by \cite{deng21}, the site boasts a median seeing of 0.75'', a median temperature variation of 2.4 degrees Celsius, and nightly precipitable water vapor consistently below 2\,mm for 55\% of the time. Being among the top observation sites in the Eastern Hemisphere, Lenghu addresses the need for a high-quality global network of observation sites for exoplanet and time-domain science.

A 9-meter retractable telescope dome is designed for the Tianyu-I and II telescopes for optimal astronomical observations, as shown in Fig. \ref{fig:dome}. Operating in temperatures from $-30$\deg to $+30$\deg, it has a 25-year lifespan. The dome includes a roof window, enabling unobstructed telescope observations, with a 360\deg\, rotation and a maximum speed of $\ge$5\deg/s. It features a self-locking mechanism, limited position switches, and computer interfaces for manual and remote control. The structure withstands rain, snow, wind, and dust, operating normally at 20\,m/s wind speeds and ensuring safety at 36.9\,m/s in the closed state. The dome's classical design incorporates advanced features, meeting the technical demands of astronomical research and providing adaptability for long-term use.

As shown in Fig. \ref{fig:infrastructure}, the two telescopes will be separated by 30\,m to guarantee unobstructed observation at an elevation angle as small as 10\deg. The foundations for the Tianyu-I and Tianyu-II telescopes support circular load-bearing domes using a circular tower structure. Each tower is around 7.8\,m tall, positioned independently from the telescope pedestals.

\subsection{Telescope, Mechanical and Control system}\label{subsec:telescope}

\begin{figure}
	\centering
	\includegraphics[width = 0.5\linewidth]{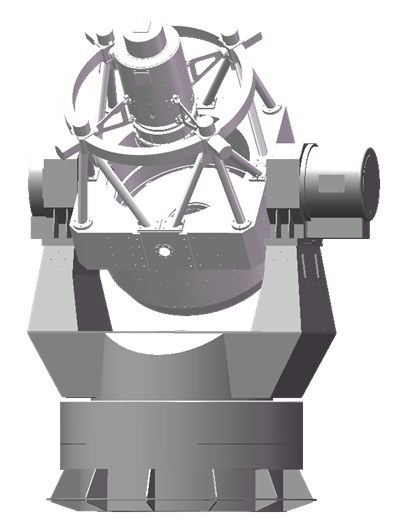}
	\caption{The views of Tianyu 1-meter telescope}
	\label{fig:telescope}
\end{figure}

\begin{table}[ht]
	\caption{Parameters of the Tianyu-I telescope}.
	\label{tab:parameters}
	\begin{center}       
		\begin{tabular}{ll} 
			\hline
			\rule[-1ex]{0pt}{3.5ex}  Parameter  & Requirement  \\
			\hline
			\rule[-1ex]{0pt}{3.5ex}  Spectral Range &	400--850 nm  \\
			\rule[-1ex]{0pt}{3.5ex}  Clear aperture 	& 1000 mm \\
			\rule[-1ex]{0pt}{3.5ex}  FOV &	4.5°  \\
			\rule[-1ex]{0pt}{3.5ex}  F-number  &	1.57   \\
			\rule[-1ex]{0pt}{3.5ex}  Focal length  &	1570 mm \\
			\rule[-1ex]{0pt}{3.5ex}  Image scale  	& 1.3"/pixel   \\
			\rule[-1ex]{0pt}{3.5ex}  Focal Plane detector	& 8Kx8K CMOS, 10$\mu$m/pixel \\
			\rule[-1ex]{0pt}{3.5ex}  Pointing accuracy  &	$<$ 5" (RMS)  \\
			\rule[-1ex]{0pt}{3.5ex}  Tracking accuracy  &	10min $\le$ 1", 10 sec$\le$0.5" \\
			\hline
		\end{tabular}
	\end{center}
\end{table}

\subsubsection{General Parameters}
Tianyu-I is a 1-meter telescope featuring a prime focus with a FOV spanning 4.5 degrees. Operating at F/1.57, it encompasses a wavelength range from 400\,nm to 850\,nm. The prime focal plane will be occupied by a 8K$\times$8K CMOS camera. Tianyu employs an altitude/azimuth mount and direct drive system which has the advantages of compact structure, simple mechanical configuration, small gravity deformation, and high pointing accuracy. The configuration of the Tianyu-I telescope is illustrated in Fig. \ref{fig:telescope}. The main subsystems of Tianyu-I are: optics system, mechanical system, and control system. The requirements for Tianyu-I are given in Table \ref{tab:parameters}.

\begin{figure}
	\centering
	\includegraphics[width = 0.6\linewidth]{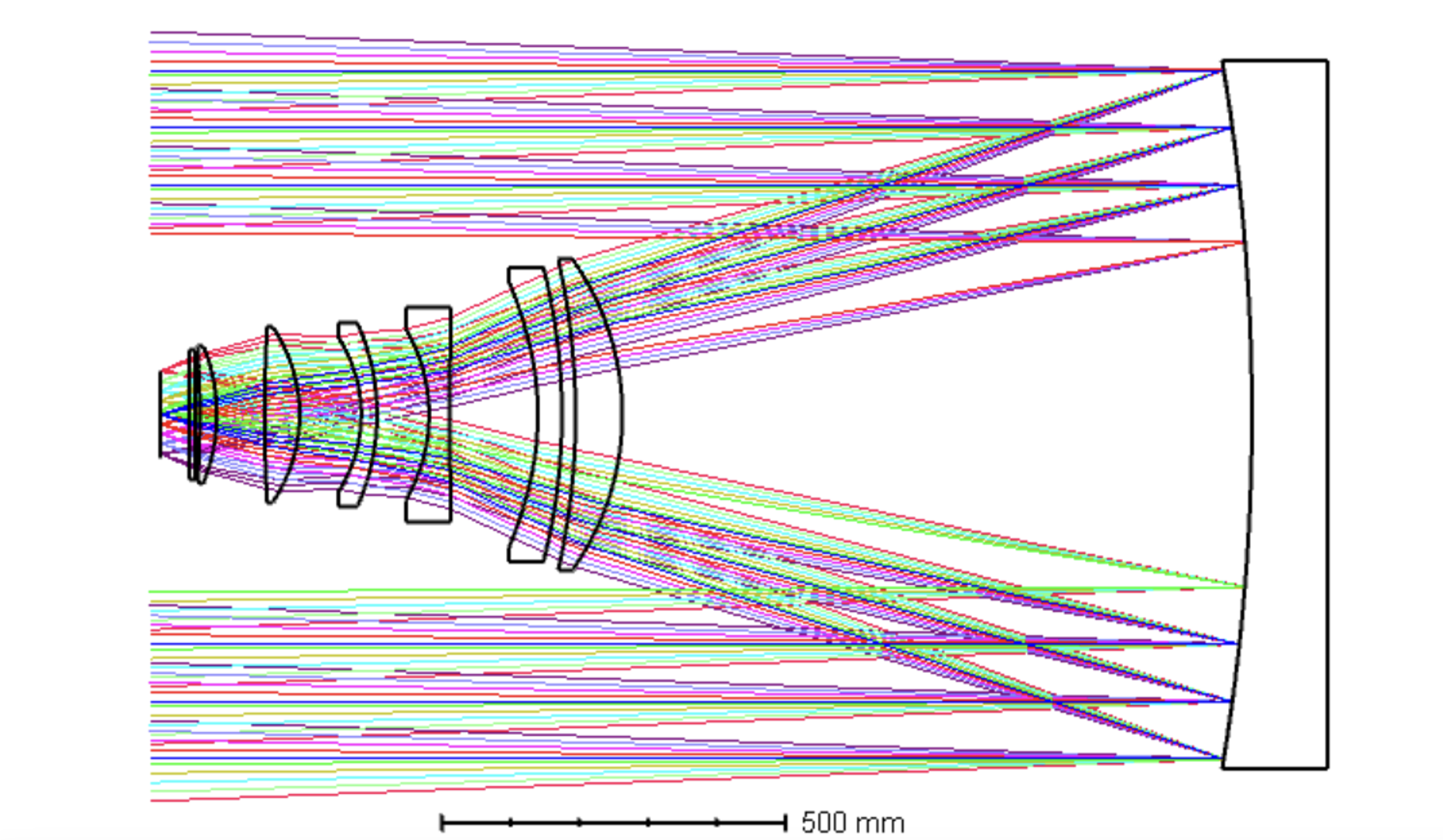}
	\caption{Layout of optical system}
	\label{fig:optics}
\end{figure}

\subsubsection{Optical system}

The optical system consists of a primary mirror with an aperture of 1\,m, 
primary focus corrector (PFC) and a filter wheel. The star light with a certain field of 
view is reflected by the primary mirror and refracted by the PFC system, converging 
and imaging on the CMOS. Optical Layout of the telescope is shown in Fig. \ref{fig:optics}. The primary mirror is a concave hyperbolic surface and material is SITALL CO-115M with nearly zero thermal coefficient of expansion.  Tianyu PFC consists of six lenses (Fig. \ref{fig:optics} ) including Fused Silica, H-FK61, and other glass material, within which one surface is aspheric. The largest lens of PFC is approximately 472\,mm in diameter, and considering the mechanical structure, the central obstruction is about 48\%. The primary mirror is coated with protected aluminum, and the PFC is coated with a multi-layer AR coating. 

\begin{figure}
	\centering
	\includegraphics[width = 0.5\linewidth]{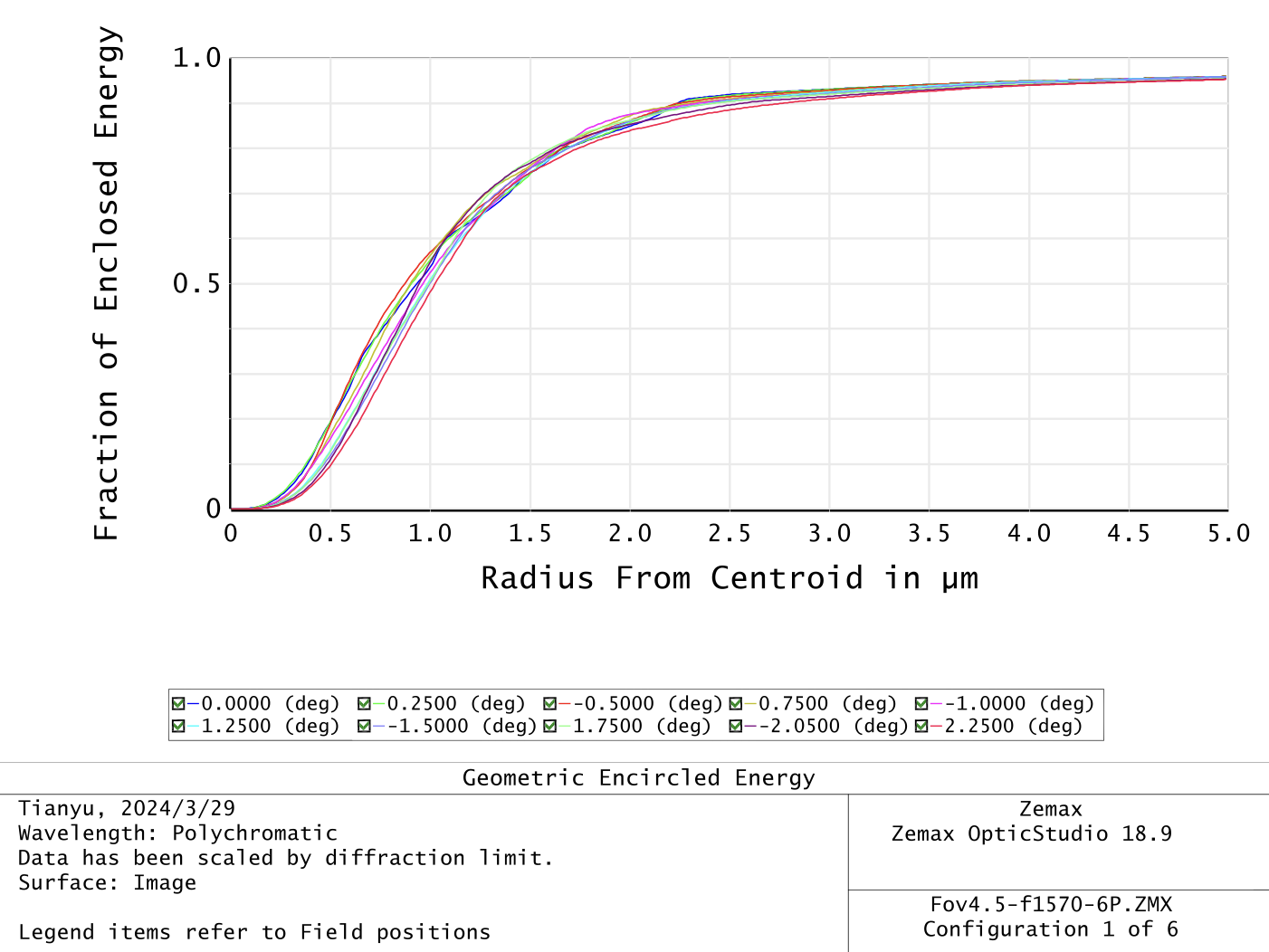}
	\caption{Geometric encircled energy diagram, The diagram shows the encircled energy 80\%(radius) less than 2\,$\mu$m, diameter less than 4\,$\mu$m or 0.5''.}
	\label{fig:EE80}
\end{figure}

\begin{figure}
	\centering
	\includegraphics[width = 0.5\linewidth]{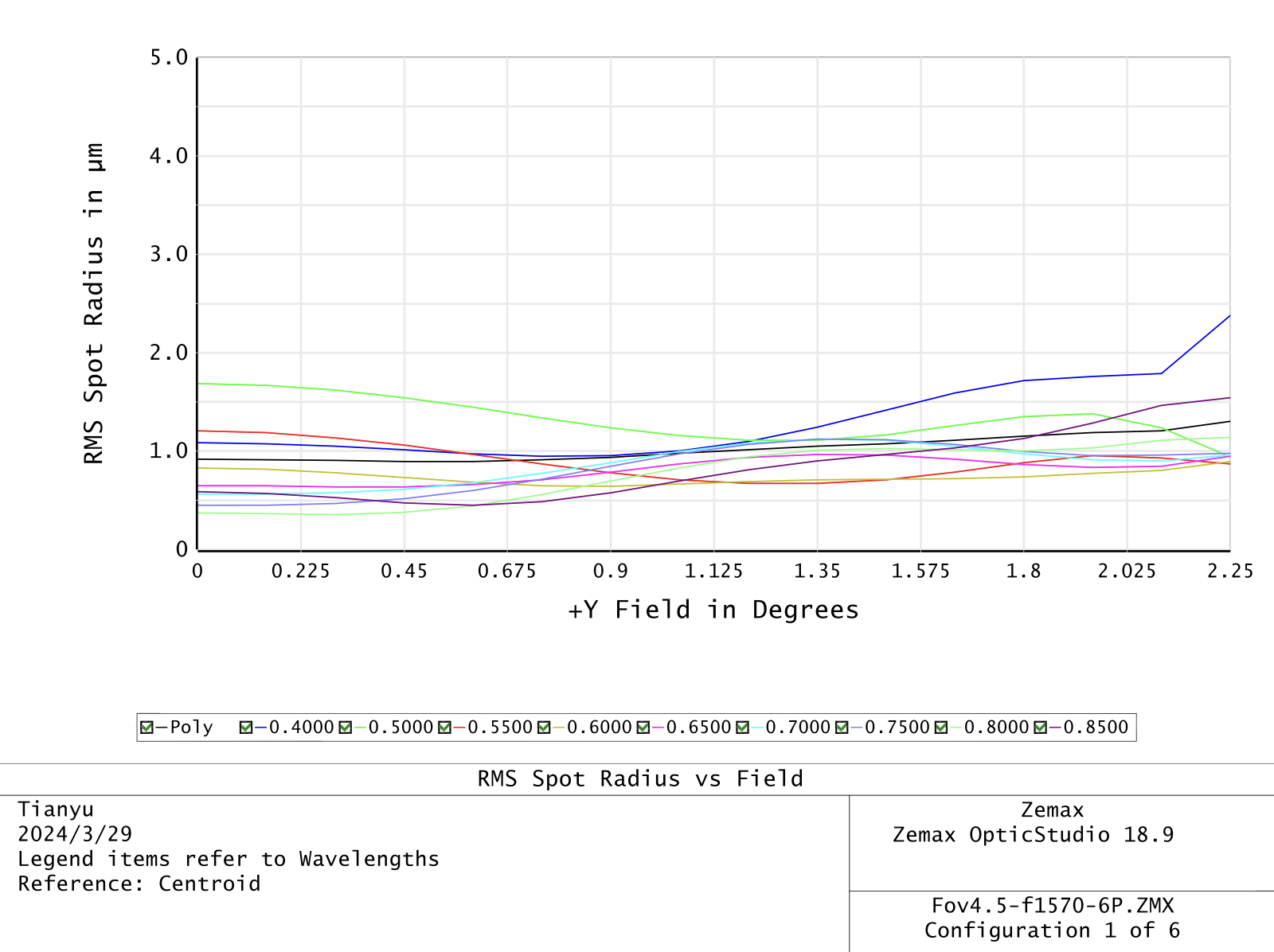}
	\caption{Variation of the RMS Spot diameter with the offset from the center of the FOV. }
	\label{fig:spot}
\end{figure}

The focal ratio is F/1.57 and it provides a resolution of 1.3''/pixel for a pixel size of 10$\mu$m. The fraction of Encircled Energy (EE) at 80\% surpasses 4$\mu$m, corresponding to an angular size of 0.5''. The RMS Spot diameter in all field are less than 5$\mu$m. The EE curves are shown in Fig. \ref{fig:EE80}, and the RMS as a function of field offset is shown in Fig. \ref{fig:spot}.

\subsubsection{Mechanical system}

The mechanical design of the telescope mainly focused on acquiring the high quality images and the requirement of accurate and stable tracking. The telescope mechanical system is composed of telescope mount, truss, top end assembly, fork, primary mirror cell, PFC pod, electric focusing mechanism, image de-rotator system, etc. (see Fig. \ref{fig:telescope}).

The telescope tracking mount is of the altitude/azimuth type. The telescope tube is a traditional open-truss Serrurier construction that can maintain the optical alignment. The primary mirror cell assembly provides a platform for the primary mirror support system. The primary mirror axial supporting system employed 18-points whiffle-tree structure. The lateral support configuration has 6 counter weight structure equally spaced around outer edge of the primary mirror. 

The PFC are installed inside a lens barrel. The lenses are separated by pads, and the material of the lens barrel is composed of steel and thermal compensation materials. The support structure of the lens contains 12 points lateral support and spacer support. The PFC pod can be assembled, tested, installed, and removed as a single unit. In order to correct temperature-induced focus changes, the whole of the PFC pod, including the camera can be displaced along the optical axis by up to $\pm$5\,mm to an accuracy of 10\,$\mu$m through electric focusing mechanism
The azimuth and elevation axis uses direct drive motors with absolute encoders. Performance requirements include a minimum rotation range of  $+/-$270\deg and the capability to position and track a point in the image at the edge of the whole FOV, achieving an RMS of less than 5''.

\subsubsection{Control system}

The telescope control system consists of the main control computer and telescope control software as the core, providing a user operation interface to achieve control functions of the telescope hardware system, including attitude control, instrument control, pointing model correction, automatic star guidance, data processing and conversion, external communication, etc. 

The azimuth and altitude axes are both directly driven by frameless torque motors, with which any mechanical drive-chain errors are eliminated. The axial position feedback sensors are Renishaw 26-bit absolute encoders. The motors are driven by a high-performance industrial 3-loop servo drive, inside which the current, velocity, and position loops are closed.

The servo drives are controlled by an EtherCAT master motion controller, with only one ethernet cable. This control method dramatically reduces the amount of wiring and enhances the reliability of data transferring.

A standard ASCOM drive is provided for the control system, with which all standard AScom client applications, such as TheSkyX, Voyage, etc., can be employed to control the telescope. Besides, self-developed control software is also provided for special usage or testing purposes.

\subsection{Instruments for Tianyu-I and Tianyu-II}\label{subsec:camera}

\begin{figure}
	\centering
	\includegraphics[width = \linewidth]{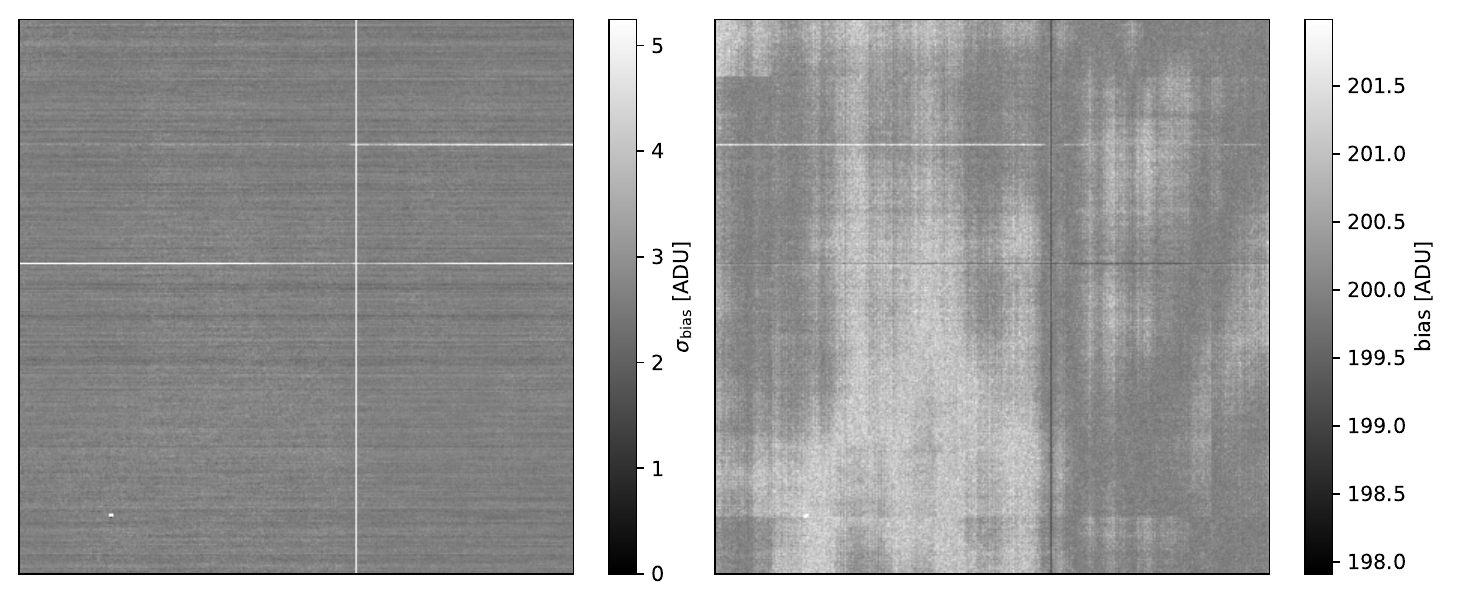}
	\caption{Left: Average bias frame for 100 frames of COSMOS66 in 18-bit HDR mode; Right: Standard deviation of bias frame for COSMOS66.}
	\label{fig:std_mean_18bit}
\end{figure}

Tianyu-I will be equipped with the COSMOS66 camera, manufactured by Teledyne Princeton Instruments\footnote{\url{https://www.princetoninstruments.com/products/cosmos-family/cosmos}}. This camera offers exceptional sensitivity, a large imaging area, low noise, and a rapid frame rate. The fundamental parameters of COSMOS66 are detailed in Table \ref{tab:cameraparam}. Utilizing the high dynamical range (HDR) mode in Tianyu-I ensures optimal photometric precision, attributed to its deep full well depth and low read noise. Fig. \ref{fig:std_mean_18bit} displays the bias frame and its standard deviation. As we see, the distribution of bias is homogeneous at the level of $<$5\,ADU.

\begin{table}
	\caption{Parameters of COSMOS66 under HDR mode. }
	\centering
	\begin{tabular}{ll}\hline
		Feature   & Parameter of COSMOS66 \\\hline
		CMOS image sensor   & Back illuminated; grade 1; 100\% fill factor\\
		Dark current at -$25^{\circ}$C&0.05 $e^-$/pixel/s\\
		Quantum efficiency& $>90$\% Peak QE\\
		Pixel format&10$\mu$m\\
		Imaging area&81mm$\times$81mm\\
		Resolution& 8120$\times$8120\\
		Full well&$\sim$119keV\\
		ADC settings&18bits\\
		System read noise&1.6$e^{-}$ rms\\
		Frame rate (FPS) & 2.9\\
		Nonlinearity &$1.61\%$\\
		Data interface&${\rm CoaXPress}^\circledR$\\
		\hline    
	\end{tabular}
	\label{tab:cameraparam}
\end{table}

Tianyu-II is envisioned as a 1-meter telescope with a substantial focal ratio falling within the range of F/6 to F/10. This telescope will be equipped with a scientific CMOS camera, a suite of multi-filters (e.g., the HB filters listed in Table \ref{tab:filters}), spectrographs, and potentially an adaptive optical (AO) system. Specifically, Sloan u', g', r', and i' filters will be employed for Tianyu-II follow-up observations. These filters are chosen due to their wide bands, enabling the efficient detection of transits, particularly for faint stars.

Our selection of spectrometers spans low, medium, and high resolutions, each serving distinct scientific purposes. The low-resolution spectrometer is tailored for supernova exploration, the medium-resolution variant supports investigations in stellar physics, and the high-resolution spectrometer is dedicated to exoplanet certification. Importantly, these cutting-edge tools will soon become accessible to astronomy enthusiasts.

\section{Survey strategy and data processing}\label{sec:strategy}
\subsection{Survey strategy}

{ In all survey modes, the exposure time for a single frame is 0.3 seconds, and these short frames would be combined into a single long frame every minute. Tianyu is designed with three survey modes that cater to the diverse requirements for scientific investigations. These modes are:}

\begin{enumerate}
	\item {\bf Short-cadence survey.} Because the short frames would be stacked to form a long frame for all survey modes, the short-cadence survey is actually embedded in both the medium-cadence and the long-cadence modes. Hence the sky coverage and number of stars of the short-cadence mode are the sum of the other two modes.     
	Tianyu-II will be used to confirm the fast phenomena that are found in Tianyu-I short frames through the alert system. This synergistic integration of Tianyu-I and Tianyu-II allow Tianyu to be sensitive to photometric variation over time scales from subsecond to hours.
	
	The primary objectives of the short-cadence survey include the detection of TNO occultations, short period variable sources, high time-resolution stellar flares, and fast optical transients. The sky coverage and number of stars in this mode is shown in Table \ref{tab:survey_strategy}. 
	
	\item {\bf Medium-cadence survey.} This survey is designed to capture 60 FOVs each night with a cadence of one hour, focusing on observing the Kepler region and TESS Northern Continuous Viewing Zone (NCVZ) with higher priority. Given that Tianyu-II will consistently monitor the intriguing phenomena initially discovered by Tianyu-I, its sensitivity will extend to events occurring within time scales ranging from 1 hour to 1 week. This aligns with the practical limit imposed by the maximum time Tianyu-II can allocate to observing a single object.
	
	Given that transiting CGs typically exhibit durations of 1-2 days, with ingress and egress lasting a few hours, a 1-hour cadence is selected as the primary sampling frequency for the medium-cadence survey. As the detection of CGs constitutes the primary scientific objective of Tianyu, approximately 80\% of the total observational time is allocated to the medium-cadence survey. Additionally, the medium-mode survey is designed to identify infant supernovae, TDEs, and NEOs.
	
	\item {\bf Long-cadence survey.} This survey is designed to capture 360 FOVs each night with a cadence of about one week, focusing on observing the ecliptic plane and the Northern sky. The long-cadence survey is sensitive to events over time scales from 1 week to 1 month through combination of Tianyu-I's alert and Tianyu-II's follow-up observations. Roughly 20\% of the total observational time is dedicated to the long-cadence survey, which aims to detect and characterise distant solar system objects,  Be stars in open cluster, and find interstellar objects. Moreover, because the long-cadence survey observe 730 million stars in about 16200 square degrees, its short frames would provide a large number of alerts, triggering Tianyu-II to find fast variables and phenomena.
\end{enumerate}
\begin{table}[]
	\caption{Summary of survey strategy. 
		Number of stars is estimated using the total area of sky region combined with the number of stars of Gaia DR3 \citep{2023A&A...674A...1G}. Because Tianyu can observe stars that is 0.5 magnitude fainter than Gaia, this is a conservative estimation.}
	\centering
	
	\begin{tabular}{llll}\hline
		Survey mode&Sensitive timescale& Total area of  sky region& Number of stars  \\\hline
		Short& 0.3 second-1 hour&18660\,$\rm deg^2$&$8.1\times10^8$\\
		Medium& 1 hour-1 week&2550\,$\rm deg^2$&$1.4\times10^8$\\
		Long& 1 week-1 month&16200\,$\rm deg^2$&$7.3\times10^8$\\\hline
	\end{tabular}
	
	\label{tab:survey_strategy}
\end{table}

\begin{figure}
	\centering
	\includegraphics[width = 0.9\linewidth]{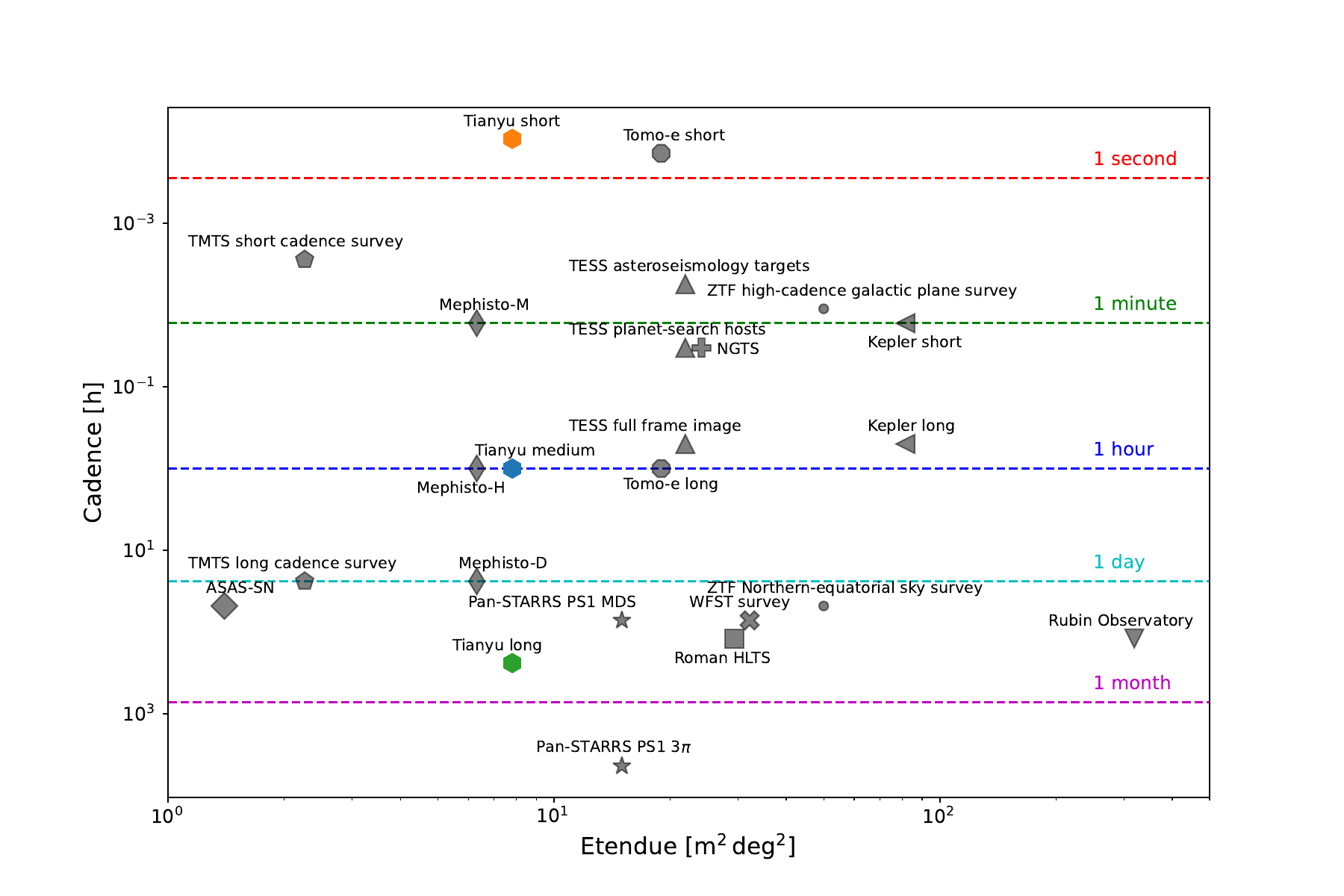}
	\caption{Cadence and etendue of Tianyu compared with other projects, including ZTF \citep{Bellm_2019}, Pan-STARRS1 \citep{Chambers2016}, TESS \citep{ricker14}, Roman \citep{Mosby2020}, Rubin observatory \citep{LSST2017}, Kepler \citep{borucki10}, WFST \citep{Chen2024}, TMTS \citep{Zhang_2020}, Mephisto \citep{Yuan2020}, ASAS-SN \citep{Hart2023}, NGTS \citep{Wheatley2017} and tomo-e \citep{sako18}), where Roman HLTS refers to Roman High Latitude Time Domain Survey; Pan-STARRS PS1 MDS refers to Pan-STARRS PS1 Medium Deep Survey.  }
	\label{fig:ca_et}
\end{figure}

Typically, cadence and etendue stand as two crucial factors for gauging the efficiency and sensitivity of a photometric survey in detecting variable sources across specific time scales. In Fig. \ref{fig:ca_et}, we compare the cadence and etendue of Tianyu with those of other photometric surveys. As evident in the figure, Tianyu stands out as a distinctive survey characterized by sub-second cadence, surpassing even the Tomo-e Gozen survey \citep{sako18}. This distinctive feature renders Tianyu particularly sensitive to extremely fast transients and variables, including but not limited to Trans-Neptunian Object (TNO) occultations, super-fast stellar flares, and other swiftly evolving phenomena. 

\begin{figure}
	\centering
	\includegraphics[width = 1.\linewidth]{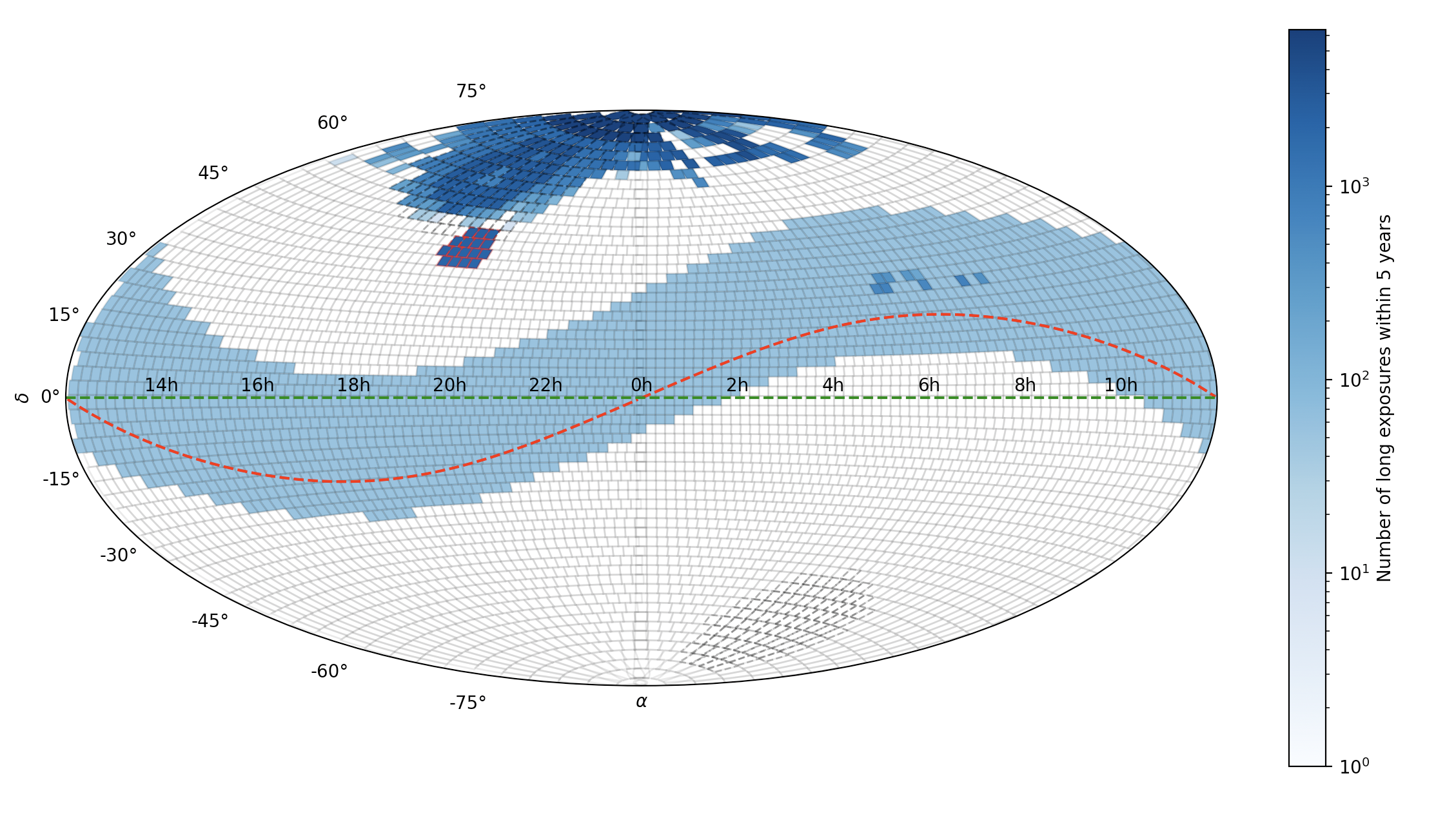}
	\caption{Footprint of Tianyu and the number of long exposures of each FOV. A long exposure is an integration of sub-second frames over one minute. The red outlined regions signify the Kepler field, while the dashed regions are observed by more than 10 TESS sectors. The darkness of the shading of the FOVs represents the number of one-minute observations over five years. The red dashed line represents the ecliptic plane. The green dashed line refers to $\delta=0^\circ$.}
	\label{fig:sky_region}
\end{figure}

The footprint of the Tianyu survey, along with the corresponding cadence, is illustrated in Fig. \ref{fig:sky_region}. In the medium-cadence survey, observations extend to the Kepler FOV and regions encompassing more than 8 TESS sectors, with the majority falling within the NCVZ. The long-cadence survey targets the region within $[-10^\circ,30^\circ]$ relative to the ecliptic plane. The short-cadence survey would focus on the raw data produced by medium-cadence survey and long-cadence survey. A summary for survey strategy is shown in Table \ref{tab:survey_strategy}. 


\subsection{Data processing}
\begin{figure}
	\centering
	\includegraphics[width = \linewidth]{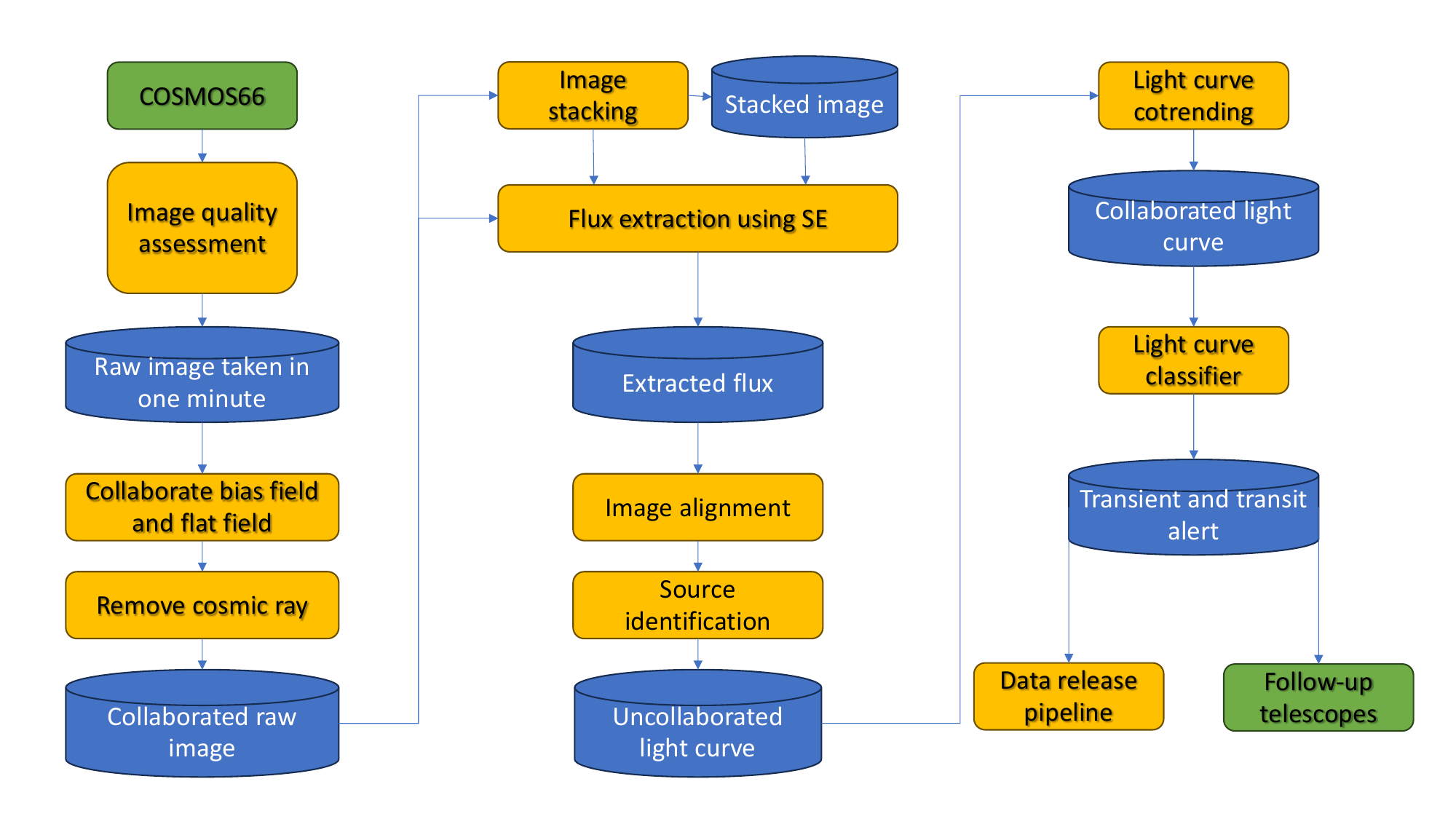}
	\caption{Workflow of Tianyu data processing. The green rectangles represent the facilities; the orange rectangles represent the pipelines; and the blue cylinders represent the data products.}
	\label{fig:workflow_DP}
\end{figure}

Two data processing and analysis frameworks are developed for the Tianyu project: the high accurate photometry and light curve classification system and the transient detection and classification system. The two systems will work together to detect and discover different kinds of time-domain signals. The workflow of the Tianyu data processing is shown in Fig. \ref{fig:workflow_DP}. The following section will discuss the details of these two frameworks.

\subsubsection{Light curve extraction}
Given the combined operation of Tianyu-I and II, real-time photometry is imperative for Tianyu-I to promptly release transient and transit alerts. These alerts are crucial for triggering timely observations by Tianyu-II and other follow-up facilities. A GPU-based stacking algorithm combined with flux extraction and image alignment using Source Extractor (SE; \citep{Bertin1996}) would enable extracting the light curve from raw images taken in one minute within 20\,s. A GPU-based stacking code using a hierarchical data structure would enable extracting flux for different time scales. The details of the pipeline will be presented in future work.

\subsubsection{Photometry and Light Curve Classification}
Our system will leverage a reinforcement learning agent as a dynamic supervisor for the photometry system. This agent automatically adjusts the parameters of the detection and photometry algorithms based on data from Gaia DR3 \citep{2023A&A...674A...1G}, ensuring optimal performance over time. Additionally, a Bayesian-based photometry neural network estimates the uncertainty associated with each magnitude measurement. Both the magnitude and its uncertainty, along with color and spectrum information, are then fed into a neural network for light curve classification \citep{cui_identifying_2023}. The classification results guide the system's actions and are also disseminated to the public science platform and expert system for further evaluation.\\

\subsubsection{Transit Detection and Classification}
For the Tianyu project, we propose a real-time transit detection pipeline that combines the strengths of machine learning algorithms with classical image subtraction. This hybrid approach leverages GPU-accelerated SE for efficient image differencing, identifying potential transient candidates. Subsequently, a dedicated transient classification algorithm analyzes these candidates by incorporating meta-information, point spread function data, and both the original and differenced images. This two-stage approach employs a macro classification framework first, followed by target-specific classifiers for refined identification. By combining machine learning's flexibility with the speed of image differencing, this pipeline promises efficient and accurate real-time transit detection.\\

\section{Photometric and timing precision}\label{sec:precision}
\subsection{Photometric precision}

Each FOV of Tianyu footprint is  allocated 60\,s per scan, utilizing stacked frames derived from the superposition of sub-frames with 0.3-second exposure.The chosen exposure strategy aims to observe stars as bright as $V=$11\,mag, ensuring that the planetary candidats around bright stars can be effectively confirmed and characterized by ground-based radial velocity facilities. 

\begin{figure}
	\centering
	\includegraphics[width = 0.9\linewidth]{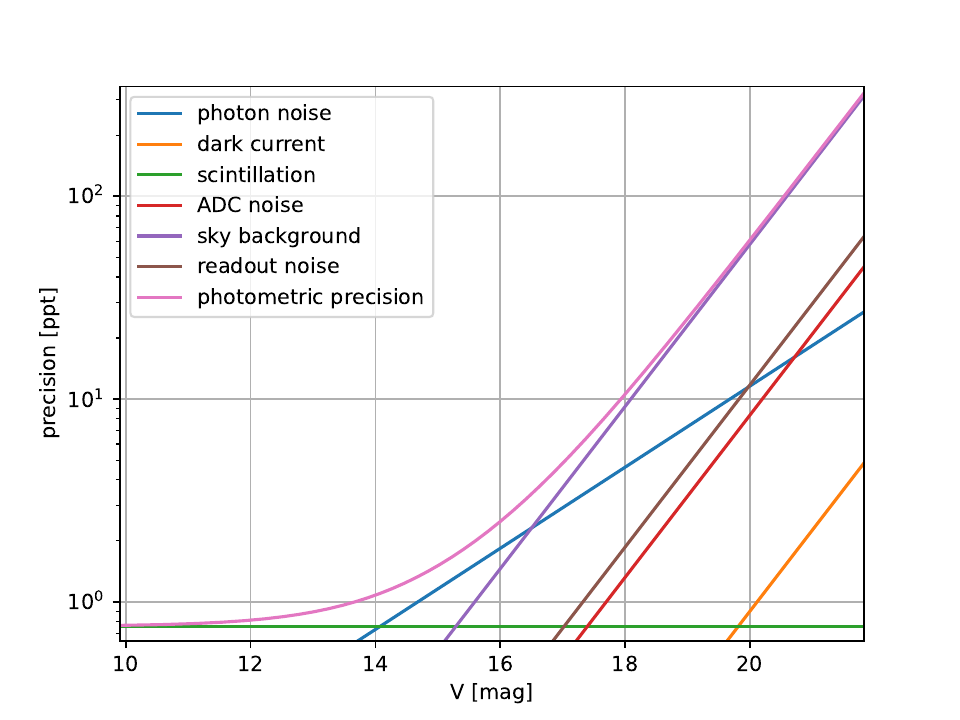}
	\caption{Photometric precision of the Tianyu medium-cadence survey for an integration time of one minute. The photometric precision is expressed in parts per thousand (ppt).}
	\label{fig:photometric_precision}
\end{figure}

The photometric precision is estimated by
\begin{equation}
	{\rm precision} = \frac{\sqrt{t_e n_{f} f_{e;\rm star} +(t_e n_{f} f_{e;\rm star} \sigma_s)^2+a_{\rm app}^2(n_{f}\sigma^2_{\rm ADC}+t_e n_{f} f_{e;\rm sky}+n_f \sigma_{e;\rm readout}^2+t_e n_f i_d)}}{t_e n_{f} f_{e;\rm star} }\,,\label{eq:photoprec}
\end{equation}
where $t_e$ is the exposure time for a single frame, $n_f$ is the number of single frames that stacked together in one minute; $f_{e;\rm star}$ is the number of photoelectron that a star would produce per unit time; $\sigma_s$ is the level of scintillation which can be estimated using Young's approximation \citep{Young1967}; $a_{app}$ is the size of aperture for photometry, which is estimated by three times of the FWHM of the PSF (including seeing); $\sigma_{\rm ADC}$ is the error caused by the quantization of ADU level; $f_{e;\rm sky}$ is the number of photoelectron that produced by the sky background within a single pixel per unit time; $\sigma_{e;\rm readout}^2$ is the readout noise; $i_d$ is the dark current. An estimation of photometric precision using eq. \ref{eq:photoprec} is shown in Fig. \ref{fig:photometric_precision}. 
Tianyu can attain a photometric precision of 0.1\% for stars brighter than $V=14$\,mag, while achieving 1\% precision for stars within the magnitude range of $14<V<18$\,mag. We expect that the photometric precision of Tianyu would be better than NGTS for bright stars \citep{Wheatley2017} and would be similar to ZTF for faint objects \citep{Masci_2019}.

\subsection{Timing precision}\label{subsec:timing_precision}
To detect high-precision TTV signals induced by non-transiting planets in a transiting system, Tianyu-II will observe transits with a sampling frequency as fine as sub-second intervals using CMOS camera with high frame rate. In this section, a simulation is undertaken to precisely measure the timing accuracy of Tianyu. The observed light curve is modeled using a template light curve $f(t)$:
\begin{equation}
	f(t) = \left\{\begin{aligned}
		&A_0+A_2(t-t_0)^2+A_4(t-t_0)^4\quad&(t_0-\frac{d}{2}\le t \le t_0+\frac{d}{2})\\
		&A_0+A_2\left(\frac{d}{2}\right)^2+A_4\left(\frac{d}{2}\right)^4\quad&{\rm otherwise},
	\end{aligned}\right. \, ,
\end{equation}
where $t_0 = 0$ is the mid-transit time; $A_0 = 0.9841796875$, $A_2 = 0$, $A_4 = 0.05$ are the coefficients to depict the shape of transit; $d = 1.5$ is the transit duration. We sample the light curve using the photometric precision shown in Fig. \ref{fig:photometric_precision}. Meanwhile, we assume stellar activity would cause white noise of 0.1\%. The uncertainty of mid-transit time is obtained using MCMC sampling. We compare the timing precision of Tianyu-II with Byrne Observatory at Sedgwick (BOS) of Las
Cumbres Observatory Global Telescope Network (LCOGT) \citep{Shporer_2010}, NGTS \citep{Wheatley2017} and superWASP \citep{2014CoSka..43..500S} for a star with $\rm V = 13$\,mag. 

\begin{figure}
	\centering
	\includegraphics[width = 0.8\linewidth]{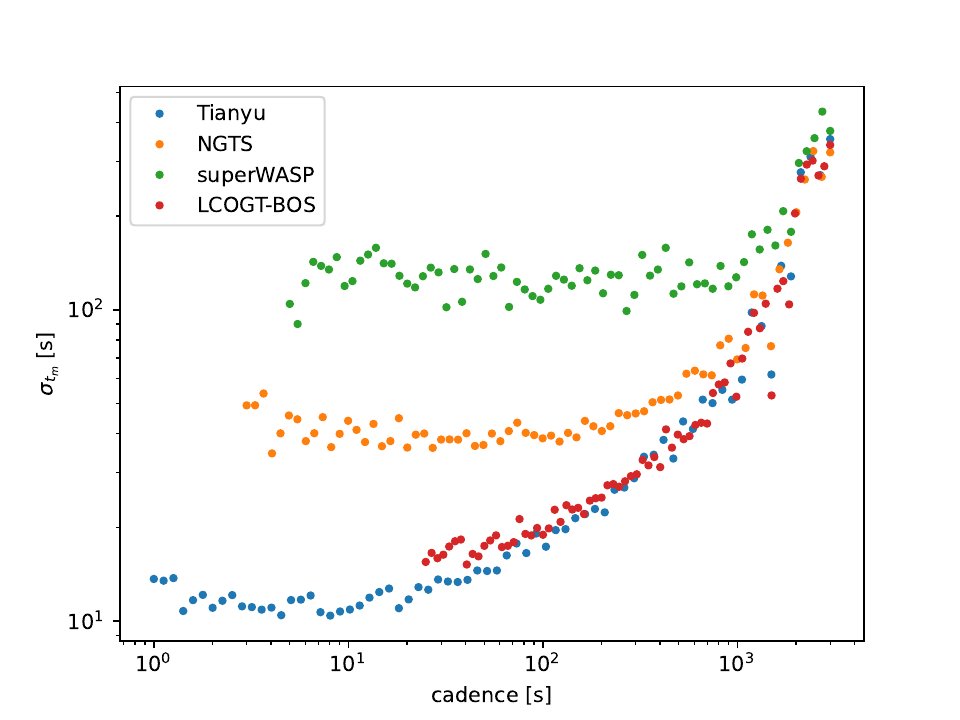}
	\caption{Timing precision of Tianyu compared with LCOGT-BOS, NGTS and superWASP. For each cadence, we use five initial observation times and calculate the median of transit-timing uncertainty.} 
	\label{fig:TTVP}
\end{figure}

Figure \ref{fig:TTVP} illustrates the timing precision of various instruments. Tianyu stands out with a remarkable timing precision of 10 seconds, achieved with a cadence of less than 10 seconds. This surpasses superWASP by more than an order of magnitude. The superior precision is credited to Tianyu's larger diameter, enhancing photometry precision compared to superWASP. Additionally, Tianyu outperforms LCOGT-BOS in timing precision, thanks to its faster readout capabilities. Figure \ref{fig:TTVP} illustrates the timing precision of Tianyu and other instruments. 

\section{Confirmation of Tianyu transit candidates}\label{sec:confirmation}

\subsection{Confirmation by Tianyu-II, Kepler, TESS and other transit surveys}\label{sec:photometry_followup}
Various astrophysical phenomena can be misinterpreted as transiting planets, especially in the case of Tianyu-I photometry without a filter or with a Luminous filter. Therefore, follow-up observations are crucial to rule out false positives and confirm transiting candidates. Numerous researchers have explored the classifications of astrophysical false positives (e.g., \citep{2003ApJ...593L.125B,2010ApJ...712...38E,2018AJ....156..234C}). Among these, eclipsing binaries, which are too close to the target to be resolved by Tianyu-I, represent a significant source of false positives. To mitigate such instances, multi-band photometry conducted by Tianyu-II will identify the color dependence of background variable sources (e.g., \citep{2022AJ....164...70Y}). Additionally, pixel-level analysis, including centroid offset analyses, difference imaging, and pixel correlation images \citep{2013PASP..125..889B}, contributes to the identification of false positives.

The identification of unresolved variable sources can also be facilitated through high-resolution imaging using AO. Additionally, indicators of contaminating binaries include differences in odd and even depths and transit shapes in the light curves collected by Tianyu-II (e.g., \citep{2012ApJ...761....6M,2018ApJ...869L...7A,2018AJ....156..234C}).

We will promptly analyze the data gathered by Tianyu-I in real-time, pinpointing transit ingress occurrences. Subsequently, we will leverage Tianyu-II to execute multi-band follow-up observations, aiming to detect single transits induced by CGs. Concurrently with Tianyu-II, we'll gather additional photometric data from missions like Kepler, TESS, and other transit surveys. This collaborative endeavor is aimed at confirming the transit candidates identified by Tianyu-I. Leveraging this collective dataset, we will validate Tianyu candidates, refine their ephemerides, and extend baselines for detecting cold giant planets and conducting TTV measurements.

\subsection{Confirmation by Gaia and JUST and other radial velocity surveys}
Given the fact that Tianyu has the great potential to reveal long-period CGs, it is essential to perform follow-up observations to confirm their real natures. As a commonly used method, the radial velocity technique is sensitive to relatively short-period and massive companions. In our solar system, the radial velocity semi-amplitude of the Sun induced by Jupiter is about $12\,{\rm m\,s^{-1}}$, feasible to be detected by high-precision and long-term stable spectrograph. Many ongoing radial velocity surveys can be dedicated to Tianyu follow-ups, such as the Automated Planet Finder (APF) at Lick Observatory \citep{2014PASP..126..359V}, NN-explore Exoplanet Investigations with Dopper spectroscopy (NEID) on Wisconsin-Indiana-Yale-NOIRLab (WIYN) 3.5-meter Telescope \citep{2016SPIE.9908E..7HS}, the Miniature Exoplanet Radial Velocity Array (MINERVA) at the Fred Lawrence Whipple Observatory \citep{2015JATIS...1b7002S} and High Accuracy Radial Velocity Planet Searcher-North (HARPS-N) on the Telescopio Nazionale Galileo (TNG) \citep{2012SPIE.8446E..1VC}. 

In addition, the upcoming JUST \citep{justteam} would provide a more superior opportunity to perform spectroscopic confirmation of objects identified by Tianyu.
JUST has a 4.4-m primary mirror, and will be equipped with a high-resolution spectrograph with an intended resolution of $R = 60,000-80,000$ and a wavelength range of 380-760 nm, making it a powerful facility to confirm the CGs that Tianyu will detect.

In addition to the radial velocity method, the Gaia astrometric data \citep{gaia16}, will be used to confirm Tianyu's findings. Gaia has measured the position of 1 billion stars with a precision as high as 20\,$\mu$as, allowing the detection of planetary-mass companions \citep{Holl2023}. Although the catalog data released by Gaia have already been used to detect and confirm hundreds of exoplanets \citep{feng22,xiao23}, Gaia's future data release (e.g. Gaia DR4; \url{https://www.cosmos.esa.int/web/gaia/release}) will provide the intermediate astrometric data (IAD) or epoch data to significantly improve the reliability and efficiency in detecting, confirming, and characterizing exoplanets. Because Gaia is especially sensitive to CGs, it will confirm the CG candidates identified by Tianyu. 

Therefore, the combination of various methods in Tianyu's follow-up strategies will enable confirmation of the transiting planet candidates identified by Tianyu and provide precise constraints of the mass and orbital parameters of these planets.

\section{Scientific yield}\label{sec:yield}
Within this section, the scientific yield is determined by incorporating the photometric and timing precision parameters discussed in section \ref{sec:precision}, along with the confirmation strategy outlined in section \ref{sec:confirmation}, for evaluating transit candidates. Initially, we will present the yield related to our primary scientific objective of identifying CGs and solar system analogs. Subsequently, we will delve into the yield associated with non-exoplanetary objects. 

\subsection{Exoplanet yield}
\label{sec:exoplanet_yield}
\begin{figure}
	\centering
	\includegraphics[width = 0.9\linewidth]{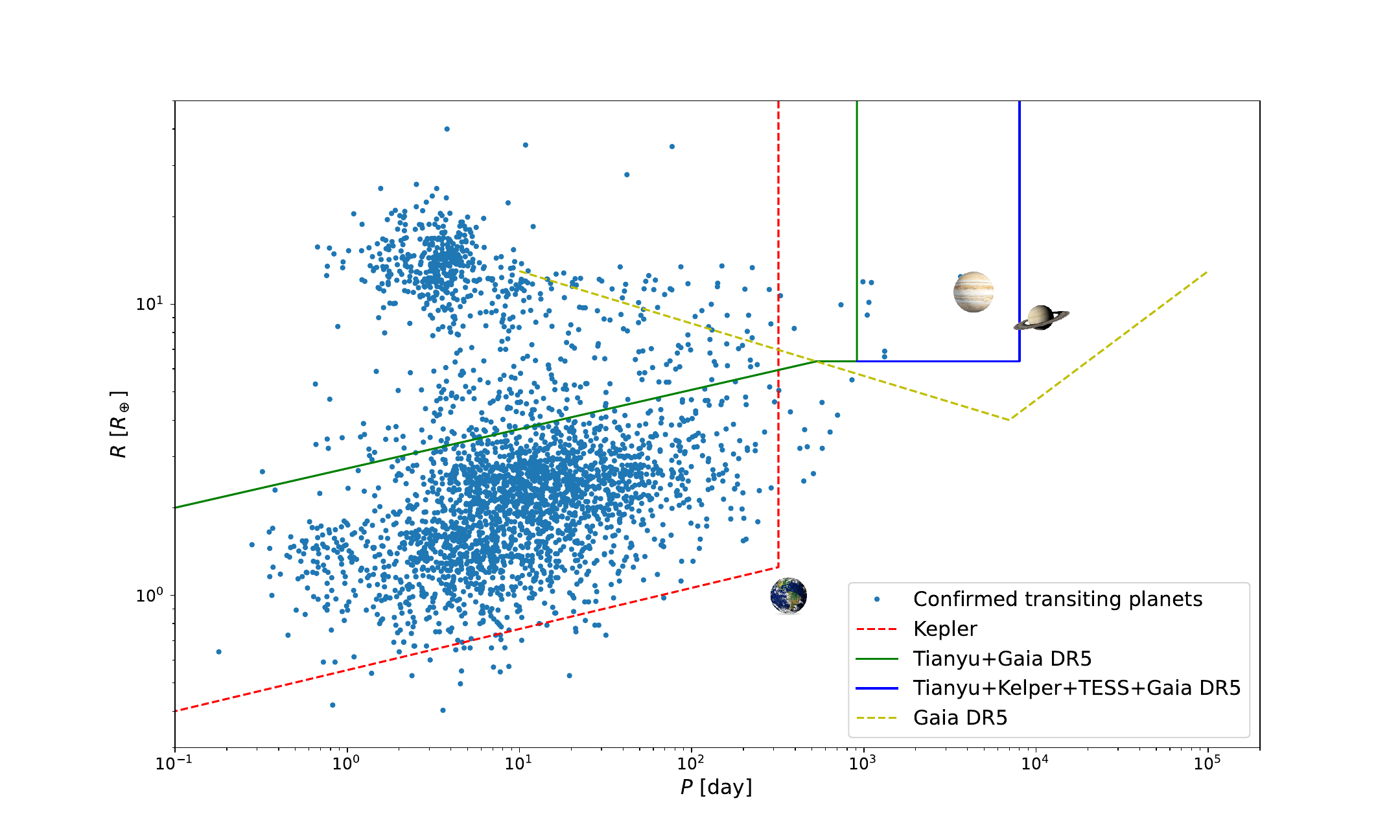}
	\caption{Detection limit of Tianyu and other surveys. Each line represents the detection limit of a survey. A transiting planet is confirmed if it is identified through at least three transits or two transits combined with Gaia astrometry.  The detection thresholds for Gaia and Kepler adhere to the methodology outlined in \cite{Sozzetti2018} and  in \cite{feng2024astrometric}, incorporating the mass-radius correlation presented by \citep{Chen_2016}. }
	\label{fig:exoplanet_param}
\end{figure}

\begin{figure}
	\centering
	\includegraphics[width = \linewidth]{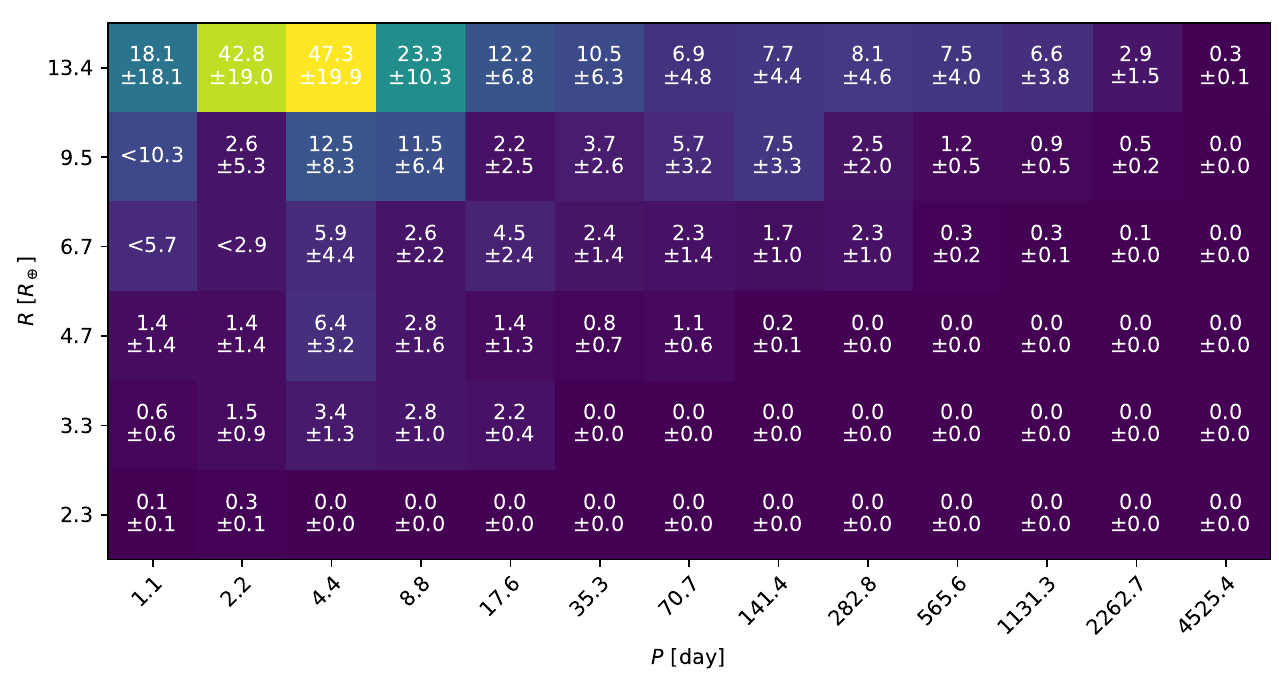}
	\caption{Estimated number of planets around FGK stars detected by Tianyu over 5 years. The occurrence rate given by \cite{Fernandes2019} is used to determine the yield. The uncertainty of the yield is derived from the uncertainty of the occurrence rate. A transit is identified when its signal-to-noise ratio is more than 5. }
	\label{fig:planetyield}
\end{figure}

Combining Kepler and TESS photometric data, Tianyu could achieve extreme-long baseline to detect Jupiter analogues, as shown in Fig. \ref{fig:exoplanet_param}. This figure illustrates the Tianyu's potential to detect CGs around solar-like stars through the use of double transits combined with astrometry. 
The turn-over region around 10 years corresponds to the transition from 3 transits to 2 transits, as determined by the combined observation baseline of Tianyu and other surveys. Detecting short-period planets is easier in this context due to the larger proportion of in-transit time among observation time, resulting in a higher signal-to-noise ratio compared to long-period planets of similar size. The yield of transiting planets around FGK stars is shown in Fig. \ref{fig:planetyield}. In this simulation, we use the criteria that the signal-to-noise ratio of the phase-folded transit should be larger than 5 to identify a detection. The signal-to-noise ratio of transit is calculated using
\begin{equation}
	{\rm SNR_{transit}}=\sqrt{N}\frac{{\rm transit\, depth}}{{\rm photometric\, precision}}\, ,
\end{equation}
where $N$ is the number of data points in a transit. We  consider the window effect in the yield prediction when simulating the signal-to-noise ratio. The observation window is set be the minimum between the typical observing time (6 hours) and transit duration. 

According to this result, 316 transiting planets would be detected by Tianyu-I. Among them, there are 12 CGs ($P>$1096\,d; $R>3 R_\oplus$ \citep{Chen_2016}), 33 warm Jupiters (20\,d$<P\le$ 365\,d; $R>10 R_\oplus$), 144 hot Jupiters ($P\le$ 20\, d; $R>10 R_\oplus$), 11 warm Neptunes (20\,d$<P\le$ 365\, d; 3$R_\oplus<R<6 R_\oplus$), and 46 hot Neptunes (20\,d$<P\le$ 365\, d; 3$R_\oplus<R<6 R_\oplus$).  

As illustrated in Figure \ref{fig:current_solar_like}, 4 CGs have been identified around solar-type stars. In contrast, the Tianyu project is expected to detect approximately 10 CGs around solar-type stars. Employing a Poisson distribution assumption, the detection of CGs by Tianyu is poised to reduce the uncertainty in the occurrence rate of CGs by a significant 47\%. 

Among FGK stars, 80\% of them are solar analogs \citep{gaia18}. The number of detected exoplanets around solar analogs can be estimated using this proportion. Assuming an occurrence rate of 10\% for Earth-like planets \citep{Fernandes2019}, the five-year survey of Tianyu will probably detect 1-2 candidates of solar system analogs, as defined in section \ref{sec:intro}. These candidates would be further confirmed by future missions searching for Earth twins, such as Plato \citep{plato14} and Earth 2.0 \citep{ge22}. 

Assuming an occurrence rate of $\sim$10\% for CGs \citep{wittenmyer20}, it is anticipated that the Tianyu project will uncover 1-2 systems featuring multiple CGs. Meanwhile, Tianyu would discover $\sim$14 systems with a CG and a hot Jupiter. Utilizing the occurrence rate in \cite{Fernandes2019}, which is $44\%$, and assuming no correlation between the occurrence rate of CGs and Neptunes, it is estimated that within a sample of 12 systems where the CGs are detected by Tianyu, approximately 5 of them would likely harbor a Neptune-sized planet  (3$R_\oplus<R<6 R_\oplus$).
Notably, these yield estimates are derived solely from Tianyu's discoveries. When integrating the findings from Tianyu with those of other transit surveys, the overall number of transiting systems harboring CGs is expected to be higher. A summary of exoplanet and planetary system yield is presented in Table \ref{tab:exoplanet_yield}. 
\begin{table}[]
	\centering
	\caption{A summary of exoplanet yield. The yield for exoplanetary systems is calculated under the assumption of coplanarity.}
	
	\begin{tabular}{lll}
		\hline
		Type of exoplanets or systems & Yield within 5 years &Reference for occurrence rate\\\hline
		CGs around FGK stars& 12& \cite{Fernandes2019}\\
		Warm Jupiters around FGK stars& 33& \cite{Fernandes2019}\\
		Hot Jupiters around FGK stars&144& \cite{Fernandes2019}\\
		Warm Neptunes around FGK stars&11& \cite{Fernandes2019}\\
		Hot Neptunes around FGK stars&46& \cite{Fernandes2019}\\
		Solar system analogs&1-2&\cite{Fernandes2019}\\
		System with multiple CGs&1-2& \cite{wittenmyer20}\\
		System with a CG and a hot Jupiter&14&\cite{wittenmyer20}\\
		System with a CG and a Neptune&5&\cite{Fernandes2019}\\\hline
	\end{tabular}
	
	\label{tab:exoplanet_yield}
\end{table}

\subsection{Yield for other scientific objectives}\label{sec:other_yield}
{With an etendue of about 25\% of ZTF and a photometric precision comparable with ZTF, it is anticipated that Tianyu's scientific output will be a quarter of ZTF's\footnote{\url{https://www.ztf.caltech.edu/}}. Compared with ZTF's 2-day nominal cadence, Tianyu's short and medium survey modes cover time scales ranging from 0.3 second to 1 week, making it more suitable for finding fast phenomena such as TNO occultation and infant supernovae. Considering these factors and according to the discussion of various scientfic objectives in section \ref{sec:goals}, Tianyu's non-planetary yield is detailed in Table \ref{tab:yield_other}. 
	
	\begin{table}[]
		\caption{Yield prediction of scientific targets. Assuming a constant occurrence rate, the yield prediction of Ultracompact binary, Supernova, Tidal disruption event and Near Earth astroid comes from the scientific results of ZTF  times the proportion of etendue between Tianyu and ZTF.}
		\centering
		
		\begin{tabular}{llll}\hline
			Object&Survey mode&Yield prediction&Section  \\\hline
			BLAP& Medium&12&\ref{subsec:pulsatingstar}\\
			HADS& Short &20 000&\ref{subsec:pulsatingstar}\\
			UCB&Short &10&\ref{subsec:EB}\\
			Be stars & Long & 250 & 2.2.4\\
			Supernova& Long&1500&\ref{subsec:SN}\\
			TDE&Short &15&\ref{subsec:TDE}\\
			NEO and comet&Medium &58&\ref{subsec:NEO}\\
			\hline
		\end{tabular}
		\label{tab:yield_other}
	\end{table}

	\section{Conclusion}\label{sec:conclusion}
	The Tianyu project, operating as a high-cadence photometric survey, is strategically focused on uncovering long-period transiting exoplanets, fast optical transients, rare variable sources, and solar system objects. Positioned at an altitude of about 4\,km in Lenghu, Qinghai, two 1-meter telescopes, Tianyu-I and Tianyu-II, are integral to this ambitious endeavor.
	
	Tianyu-I boasts a substantial FOV and is equipped with a CMOS large-format camera, ensuring efficient signal detection. On the other hand, Tianyu-II is slated to feature multi-band filters, spectrographs, and potentially adaptive optics for signal confirmation and classification. To optimize scientific yields, Tianyu will allocate 80\% of its total time to a 1-hour medium-cadence survey, with the remaining 10\% each dedicated to high and low-cadence surveys.
	
	The medium-cadence survey aims for exceptional photometric precision—less than 0.1\% for stars with magnitudes between 11 and 14, and less than 1\% for stars with magnitudes between 14 and 18. With a magnitude limit of approximately $V=21$, comparable to ZTF's transient detection limit, Tianyu-I is poised to discover over 300 transiting exoplanet candidates, including about 12 cold giants. Confirmation will be conducted by Tianyu-II, Gaia, JUST, and other facilities, providing crucial insights into the formation and evolution of giant planets.
	
	The survey's capabilities extend to detecting early phases of supernovae, tidal disruption events, kinonova, and electromagnetic counterparts of gravitational events detected by LIGO, VIRGO, and KAGRA. Additionally, Tianyu will conduct short-cadence surveys for open clusters, aiming to uncover short-period transiting exoplanets and variable sources. The low-cadence survey is dedicated to detecting distant solar system objects such as interstellar objects and TNOs. Tianyu data will contribute to the detection and characterization of NEOs, MBAs, TNOs, and active small bodies.
	
	Beyond its scientific mission, Tianyu serves as a platform for public engagement through data sharing and outreach events. The development of data reduction and analysis pipelines is underway, with the anticipated first light in 2025. Uniquely designed for both exoplanet detection and time-domain science, Tianyu's observational strategy leverages existing data from TESS, Kepler, and other surveys, enhancing sensitivity to long-period transiting planets. Equipped with a high frame-rate CMOS detector, Tianyu is poised to excel in discovering fast optical transients, potentially unveiling a new class of these phenomena.
	
	\section*{Acknowledgements}
	We would like to thank Rob Wittenmyer and Pablo Peña for helpful discussions. We extend our sincere appreciation to Ensi Liang of Teledyne for generously providing the test data for the Tianyu-I CMOS camera. Additionally, we express our gratitude to the anonymous referee for their valuable comments, which have greatly contributed to the enhancement of this manuscript. We express our gratitude to Shanghai Jiao Tong University for their invaluable support in the construction of the Tianyu-I telescope. Additionally, we extend our appreciation to the Qinghai provincial government and Haixi prefecture for preparing the site and necessary infrastructure.
	
	This work is supported by Shanghai Jiao Tong University 2030 Initiative, Science and Technology Commission of Shanghai Municipality  (project No. 23JC1410200) and Zhangjiang National Innovation Demonstration Zone  (project No. ZJ2023-ZD-003). JSJ is supported by the China-Chile Joint Research Fund under project CCJRF 2205, by FONDECYT grant 1201371, and from the ANID BASAL project FB210003.
	Y.-Z. Cai is supported by the National Natural Science Foundation of China (NSFC, Grant No. 12303054), the Yunnan Fundamental Research Projects (Grant No. 202401AU070063) and the International Centre of Supernovae, Yunnan Key Laboratory (No. 202302AN360001). Ming Yang is supported by the NSFC (Grant No. 12150009).
	Xiaoying Pang is supported by the National Natural Science Foundation of China through grants 12173029 and 12233013.
	Wenxiong Li is supported by the National Natural Science Foundation of China (NSFC Grant No. 12120101003 and 12233008). Xian Shi is supported by the National Natural Science Foundation of China (NSFC Grant No. 12233003).






\bibliographystyle{aasjournal}
\bibliography{tianyu}



\end{document}